\documentclass[a4paper,11pt]{article}

\usepackage{geometry}
\usepackage{times}
\usepackage{graphicx}
\usepackage{amsmath,amssymb}
\usepackage{authblk}
\usepackage[hidelinks]{hyperref}
\usepackage{multicol}
\title{Design and characterization of the POKERINO prototype for the POKER/NA64 experiment at CERN}
\author[1]{Andrei Antonov}
\author[1]{Pietro Bisio}
\author[2]{Mariangela Bondì}
\author[1,3]{Andrea Celentano}
\author[1]{Anna Marini\thanks{Corresponding author: anna.marini@ge.infn.it}}
\author[1]{Luca Marsicano}
\affil[1]{ \quad Istituto Nazionale di Fisica Nucleare, Sezione di Genova}
\affil[2]{\quad Istituto Nazionale di Fisica Nucleare, Sezione di Catania}
\affil[3]{\quad Universit\`a degli studi di Genova, Dipartimento di Fisica}

\date{\today}

\begin{document}

\maketitle

\begin{abstract}
The NA64 experiment at the CERN H4 beamline recently started a high-energy positron-beam program to search for light dark matter particles through a thick-target, missing-energy measurement. To fulfil the energy resolution requirement of the physics measurement $\sigma_E/E\simeq2.5\%/\sqrt{E\mathrm{[GeV}]} \oplus 0.5\%$ and cope with the constraints and performance requests of the NA64 setup, a new high-resolution homogeneous electromagnetic calorimeter PKR-CAL has been designed. The detector is based on PbWO$_4$ crystals, each read by multiple SiPM sensors to maximize the light collection. The PKR-CAL design has been optimized to mitigate and control unavoidable SiPM saturation effects at high light levels, as well as to minimize the gain fluctuations induced by instantaneous variations of the H4 beam intensity. The $R\&D$ program culminated in the construction of a small-scale prototype, POKERINO. In this work, we present the results from the experimental characterization campaign of the POKERINO aiming at demonstrating that the obtained performances are compatible with the application requirements.
\end{abstract}

\begin{multicols}{2}

\section{Introduction}


The existence of dark matter (DM) is supported by numerous astrophysical observations — indicating that it makes up about 85$\%$ of the total mass of the Universe. Still, the microscopic nature of this new form of matter  remains completely unknown~\cite{Bertone:2004pz}. 
Amidst the various theoretical models, the ``light dark matter'' (LDM) hypothesis assumes that DM consists of particles with a mass similar to or smaller than that of the proton, interacting with ordinary matter through a new force. More generally, the so-called ``Dark Sector'' scenario assumes that LDM is the lightest stable state of a new sector in Nature, with its own particles and forces. 
Among the different experimental techniques adopted to probe the LDM hypothesis~\cite{Alexander:2016aln,Battaglieri:2017aum,Filippi:2020kii,Fabbrichesi:2020wbt,Graham:2021ggy}, the \textit{missing-energy} approach proved especially effective, setting the most stringent exclusion limits in vast regions of the LDM parameters space. In this scheme, a medium energy  ($\sim$ 10$\div$100 GeV) electron or positron beam impinges on a thick active-target (an electromagnetic calorimeter), which is used as a detector to measure the energy deposited by each impinging particle. If LDM is produced in the thick-target by the interaction of the beam with its constituents, it carries away a significant fraction of the primary beam energy. The expected signal is thus the observation of events characterized by a large ``missing energy'' $E_{miss}$, defined as the difference between the nominal beam energy $E_{0}$ and the energy measured by the calorimeter $E_{cal}$. 

The only currently operating experiment exploiting the \textit{missing-energy} technique is NA64$-e$ at CERN, running a comprehensive experimental program to search for LDM, in a fixed-target electron-beam setup~\cite{PhysRevLett.131.161801} at the North Area H4 beamline. 
In parallel to the main electron-beam effort, in 2022 NA64 started a positron-beam experimental program, supported by the POKER ERC project~\cite{Marsicano:2024aR}. 
The goal of the positron-based measurement is to exploit a recently-proposed LDM production mechanism based on resonant $e^+$ annihilation on atomic electrons~\cite{PhysRevLett.121.041802}. The main advantage of the new approach is that signal events manifest through a very distinct dependence on the $E_{miss}$ observable, distributed according to a narrow peak whose position solely depends on the LDM force carrier mass~\cite{PhysRevD.104.L091701}. In order to exploit this peculiar signal signature to enhance the experimental sensitivity, the POKER project proposed to implement a new, high-resolution PbWO$_4$ calorimeter PKR-CAL, to be used as an active target for the NA64 positron program. 
The positron-beam missing-energy measurement poses demanding performance requirements for the PKR-CAL. The detector should  measure individually the energy deposited by each particle therein, while properly disentangling signals that are close in time. It should fully contain the electromagnetic shower initiated by each high-energy positron, and provide sufficient energy resolution to accurately reconstruct the expected signal peak in the missing-energy distribution. In addition, the detector should be compact enough to allow for its installation within the NA64-$e$ experimental setup,  in a maximum length along the beam line of $\sim$50 cm.  A dedicated Monte Carlo study based on the DMG4 package for simulating LDM production in $e^+e^-$ annihilation~\cite{Oberhauser:2024ozf} indicates that, in order to resolve even the narrowest predicted signal peaks, the calorimeter must achieve an energy resolution of approximately $\sigma_E / E = 2.5\% / \sqrt{E\mathrm{ [GeV]}} \oplus 0.5 \%$.  Such performance has to be achieved in the measurement conditions, where the PKR-CAL is exposed to a 100 GeV/c beam with a typical particle rate $f_{b}$ equal to 100 kHz. 
Due to the specific production mechanism of the electron/positron beams at H4, 
particles impinge on the detector randomly in time, 
and short-term instantaneous variations up to $\approx 30\%$ are possible, also depending on the overall North Area beams configuration~\cite{Papotti:2023gsn}.


The calorimeter design devised to meet the specifications consists in a 25 radiation lengths homogeneous detector, made of a $9\times9$ matrix of $20\times20\times220$ $ $mm$^3$ PbWO$_4$ crystals, with a $\sim4$ radiation lengths pre-shower section. 
The energy resolution and response time specification, as well as compactness requirements, make SiPMs as the optimal choice for the light readout. To our knowledge, this is one of the first examples of a homogeneous electromagnetic calorimeter read-out by SiPMs for high-energy (10-100 GeV) physics, another notable example being the Crilin semi-homogeneous detector for the future muon collider~\cite{Ceravolo:2022rag}, longitudinally segmented into multiple $\approx 4$ $X_0$ layers. 
These considerations motivated a dedicated R$\&$D effort and extensive preliminary testing. 
To validate the design and technical choices of the PKR-CAL, and in particular to prove the performance of the SiPMs-based light-readout system in the demanding operation condition of the planned PKR-CAL measurement, we constructed a small-scale prototype of the calorimeter, the POKERINO detector, composed by a $3\times3$ matrix of  PbWO$_4$ crystals. 
In this work, 
we present the results from the experimental characterization campaign of POKERINO aiming at demonstrating that the obtained performances are compatible with the application requirements. Specifically, we focus on the results obtained from the POKERINO characterization with cosmic rays and a test-beam at the CERN H6 beam line during summer 2024. 

\section{Materials and Methods}
\subsection{The POKERINO prototype}

The POKERINO detector design reproduces, on a small scale, all the main technical details of the full PKR-CAL detector. The prototype is a $3\times3$ matrix of $2\times2\times22$ cm$^3$ PbWO$_4$ crystals. Each crystal is wrapped with a VM2000 adhesive foil (Specular Film DF2000MA from 3M, nominal thickness 100~$\mu$m), and further wrapped with a black Tedlar foil to prevent light cross-talk among adjacent samples. 
The scintillation light produced in each crystal is read out by an assembly of four $6\times6$~mm$^2$ SiPMs with a pixel size of 10 $\mu$m (model S14160-6010 from Hamamatsu, see also Tab.~\ref{tab:SiPM}) - further details on the SiPM assembly, hereafter referred to as \texttt{PKR-CAL-SiPM}, are given in the next section. The SiPMs are optically coupled to one of the crystal end faces using RTV 3145 optical glue, with a VM2000 reflective adhesive leaflet attached to the opposite crystal face to enhance light collection efficiency. With this configuration, approximately $70\%$ of the crystal face is equipped with readout sensors. Given the nominal PbWO$_4$ light emission yield of about 100 photons/MeV, an overall light yield of about 5 phe/MeV is expected, well matched to the energy resolution requirement. 

\begin{figure*}
    \centering
    \includegraphics[width=.45\textwidth]{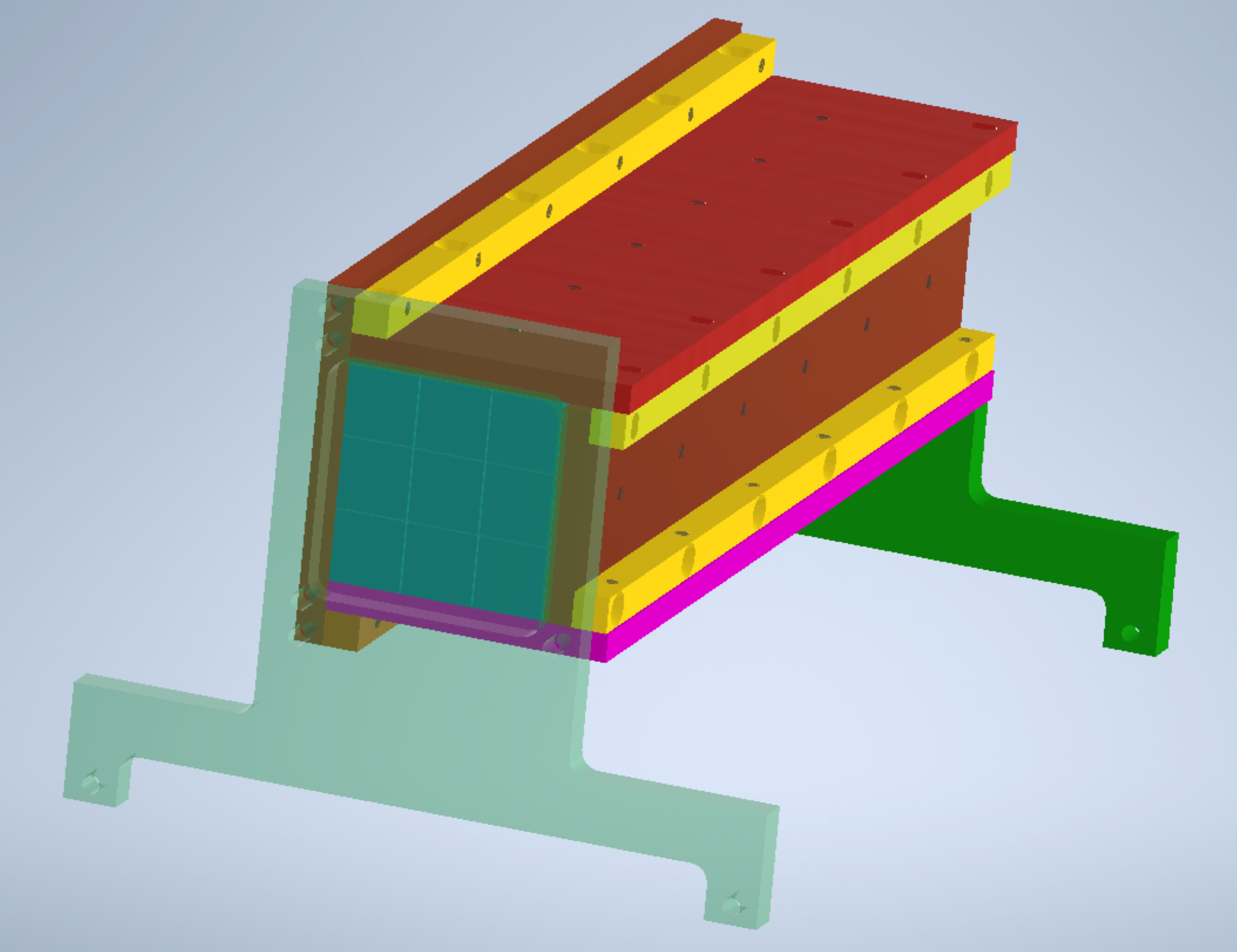}
    \quad
    \includegraphics[width=.45\textwidth]{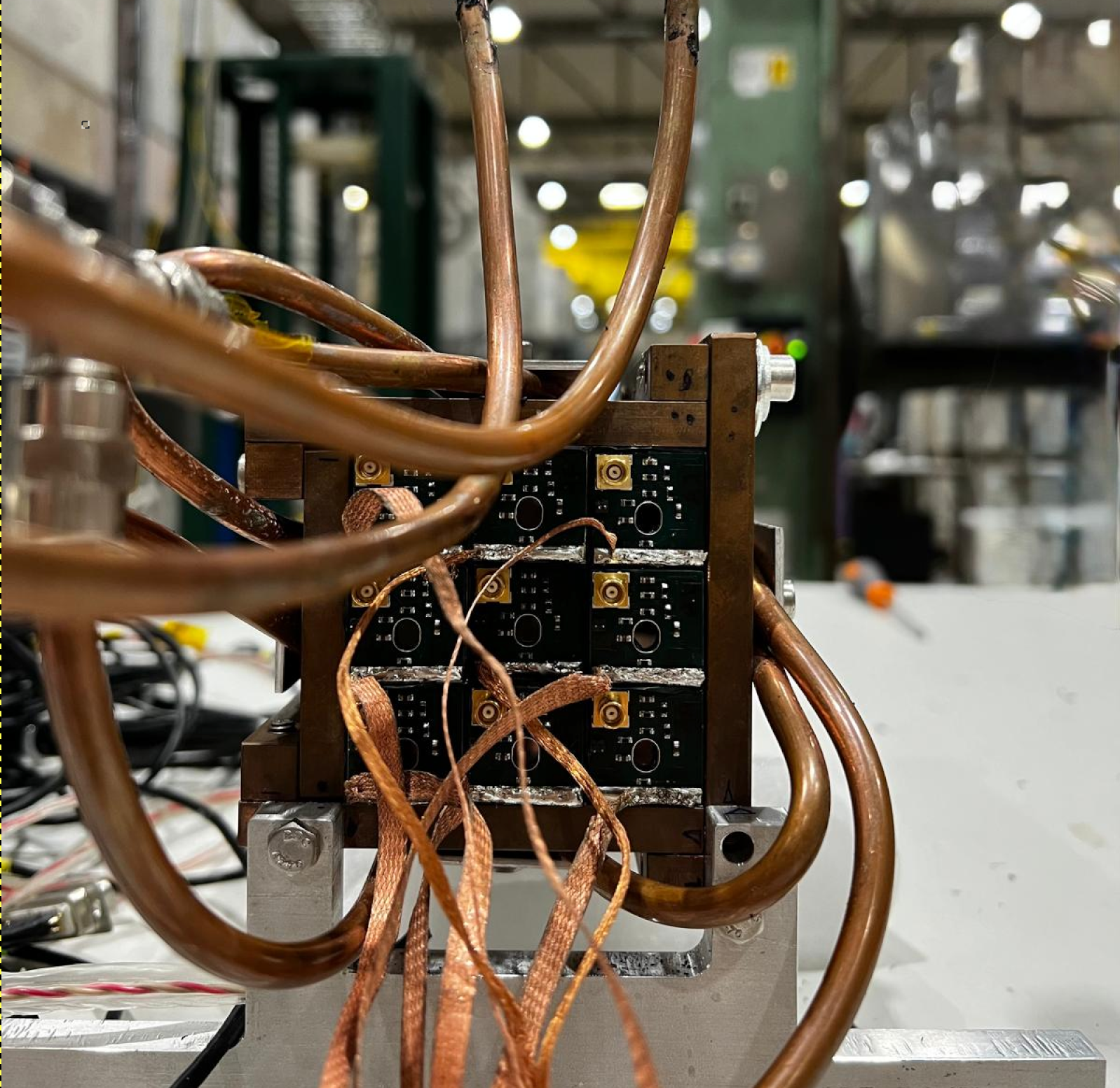}\\
    \caption{(\textbf{a}) CAD drawing of the POKERINO mechanical assembly. The PbWO$_4$ crystals (in cyan) are inserted in the copper-based mechanical assembly, formed by four plates (brownish) kept together by copper bars (yellow). The prototype is kept in position by two lightweight aluminium supports (green). The PKR-CAL-SiPM sensors are glued on the opposite crystals faces. \textbf{b}) A picture of the POKERINO detector, seen from the readout side. The copper braids are used to thermally couple the \texttt{PKR-CAL-SiPM} devices to the external copper structure.}
    \label{fig:ProtoMecc}
\end{figure*}
\begin{table*}[b]
\centering
\caption{Hamamatsu S14160-6010 main properties, where $V_{BD}$ is the device breakdown voltage.\label{tab:SiPM}}
\label{tab:SiPM}
\begin{tabular}{lll}
{Parameter} & {Conditions} & {Value} \\
\hline
Active Area  & & 6 $\times$ 6 mm$^2$\\
Number of pixels &  & 359011\\
Quantum Efficiency & $\lambda=460$ nm& 18$\%$ \\
Detector Capacitance $C$ & $V_{b}=V_{BD}+5V$ & 2.2 nF \\
Intrinsic Gain & $V_{b}=V_{BD}+5V$ & 1.7$\cdot 10^{5}$\\
\hline 
\end{tabular}
\end{table*}

The nine crystals are inserted in a mechanical structure made of 4 copper plates, connected to each other with screws. The mechanical assembly is held by two aluminium supports, with a 2-mm thickness for the part in front of the crystals, where the particles impinge. A CAD drawing of the detector is reported in Fig.~\ref{fig:ProtoMecc}, together with a picture of the real device.
In order to minimize the dead materials in front of the crystals, no further support structures are present. Exploiting the mechanical structure, crystals alignment is obtained by mounting the first two copper plates (bottom and right) in a ``L''-like structure, and then assembling  the crystal matrix on it. Finally, the mechanical structure is completed with the left and the top copper plates, exerting a minimal pressure on the crystals while closing the screws. Two copper cooling pipes (not shown in the CAD drawing) are welded, respectively, on two of the copper plates (top/right and bottom/left) and connected to an external water chiller (Lauda RP PRO 245E) to provide temperature stabilization with $\pm$0.1 $^\circ$C. 
Multiple Pt100 sensors are coupled to the mechanical structure, as well as to the inlet and outlet cooling pipes, to monitor the detector temperature. Finally, a 1-cm thick 2x2 plastic scintillator tile, read by a SiPM, is mounted on the front aluminium frame, in front of the centremost crystals, to allow for a simple centring of the beam on the detector. The detector is enclosed in a light-tight black box, whose front face has a hole matched in size and position to the crystals. A thin Tedlar foil attached with Kapton adhesive tape covers the hole ensuring the light-tightness of the whole assembly. Coaxial cables from the \texttt{PKR-CAL-SiPM} boards are routed out of the box via feed-through coaxial LEMO connectors mounted on the top face of the box. The Pt100 sensors are read in a four-wire configuration. Wires from the sensors are soldered on the pins of a DB 25-connector also mounted on the top face, to allow the connection with an external measurement system (OMEGA CYD 218E). 

The \texttt{PKR-CAL-SiPM} sensors and the SiPM reading the plastic scintillator tile were connected to ten PKR-CAL trans-impedance amplifiers, with nominal current-to-voltage gain of $50\,\Omega$. The sensors' bias voltage was provided by a Wiener MPOD 8060 board, with a custom voltage distribution circuit developed to connect a single HV board to all amplifiers. The amplifiers' output signal was digitized by a custom 14 bits, 250 MS/s ``Waveboard'' ADC~\cite{AMELI2019286}. A reduced version of the NA64 DAQ system was used during the measurements~\cite{Salamatin:2023loh}. Data was reconstructed through the standard NA64 software providing, for each event, the amplitude in ADC units and the time relative to the trigger time of the recorded pulses. 

\subsubsection{The \texttt{PKR-CAL-SiPM} sensor}

\begin{figure*}[t]
    \centering
    \includegraphics[width=0.5\linewidth]{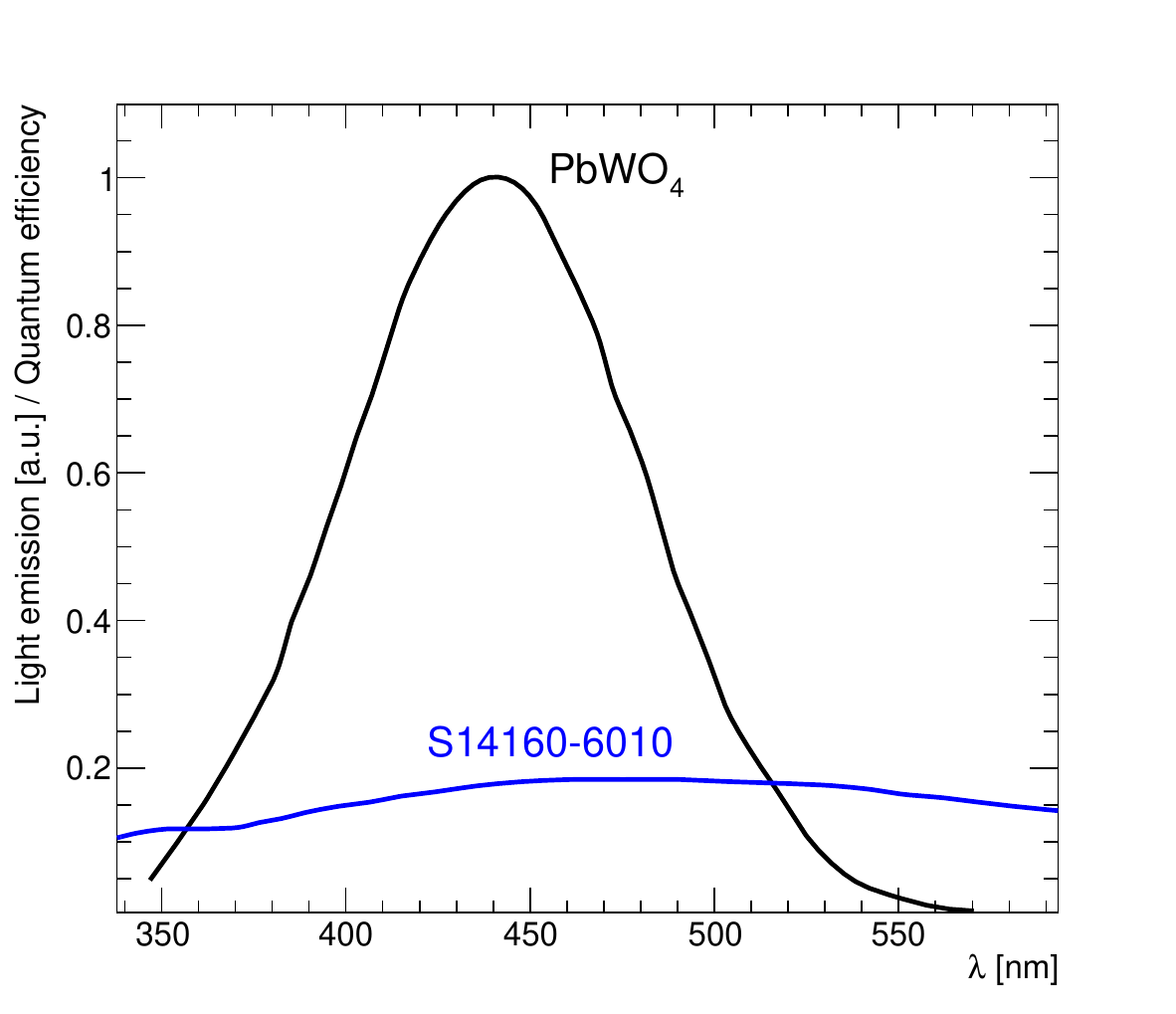}
    \caption{The typical light emission spectrum of a PbWO$_4$ crystal (black), compared with the quantum efficiency of the Hamamatsu S14160-6010 SiPM (blue) for a bias voltage of $V_{BD}+5$~V.}
    \label{fig:1}
\end{figure*}

\begin{figure*}[t]
    \centering
    \includegraphics[width=.48\textwidth]{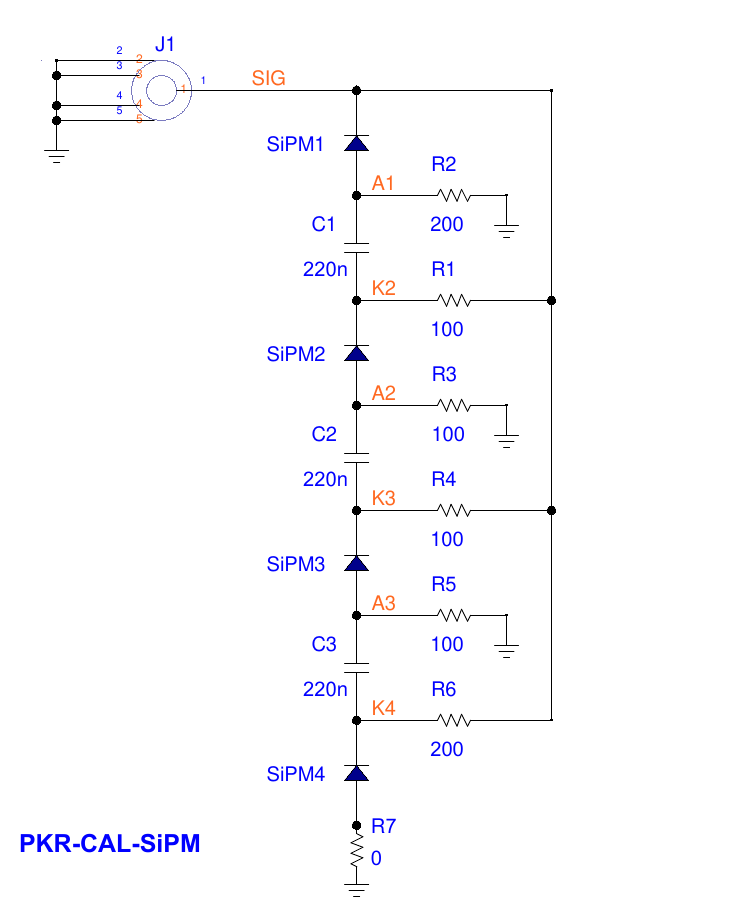}
    \caption{Schematic of the \texttt{PKR-CAL-SiPM} assembly, with four SiPM detectors connected together in the hybrid configuration. Close to each capacitor (resistor) the corresponding value in Farad (Ohm) is reported.}
    \label{fig:PKR-CAL-SiPM1}
\end{figure*}

In the PKR-CAL detector the 4 photosensors of each \texttt{PKR-CAL-SiPM} assembly are connected together using a novel hybrid approach, already tested by the MEG experiment~\cite{Ieki:2018pbf}, to minimize the readout channels and the high-voltage connection, as well as to allow operations at the nominal bias voltage of each individual SiPM.
The connection scheme is shown in Fig.~\ref{fig:PKR-CAL-SiPM1}. The ``SIG'' input is connected, through a few-m long 50 $\Omega$ coaxial cable, to a signal amplifier board, that also provides the bias voltage for the sensors. In this scheme, the capacitors, whose value is significantly larger than the intrinsic SiPM capacitance, act as an open circuit for the bias voltage source, operating in the DC regime, and thus the four SiPMs are actually provided with the same bias voltage by the resistors network, as if they were connected in \textit{parallel}. For the fast pulses induced by SiPM pixels discharge, instead, the capacitors act as a short circuit, and thus the readout amplifier sees an equivalent \textit{series} combination of the devices.

The value of the bias resistors $R_1 \ldots R_6$ in the circuit was carefully selected to fulfil the following requirements. First, it should be high enough to guarantee that the current pulse induced by SiPM discharge following a light flash propagates through the capacitor branch of the circuit - at first approximation, this corresponds to the requirement $R \gg (\omega C)^{-1}$, where $\omega = 2\pi f \simeq 2\pi/\tau$, with $\tau$ being the typical SiPM response time. Since $\tau \simeq$~10~ns, this corresponds to $R \gg 1\,\Omega$. At the same time, the resistors value should be kept as low as possible to avoid a large bias voltage drop due to the DC current flowing through them. Other than generating local heat that has to be properly dissipated to avoid temperature (and thus gain) variation of the photosensors, a large DC current could affect the SiPM response in case of any short-term variation of the beam intensity, due to the corresponding large variation on the voltage drop on the resistors. The selected values are a trade-off between these two requirements.

\begin{figure*}[t]
    \centering
    \includegraphics[width=.3\textwidth]{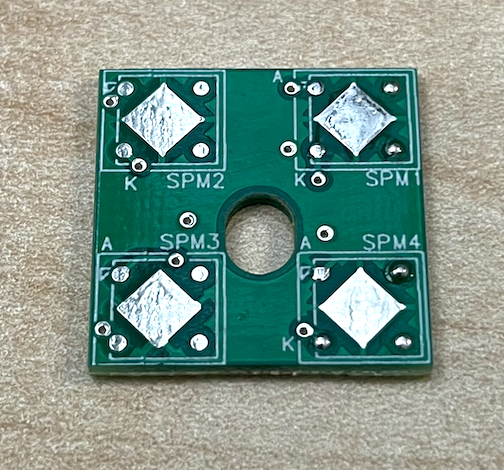}
    \quad
    \quad
    \includegraphics[width=.28\textwidth]{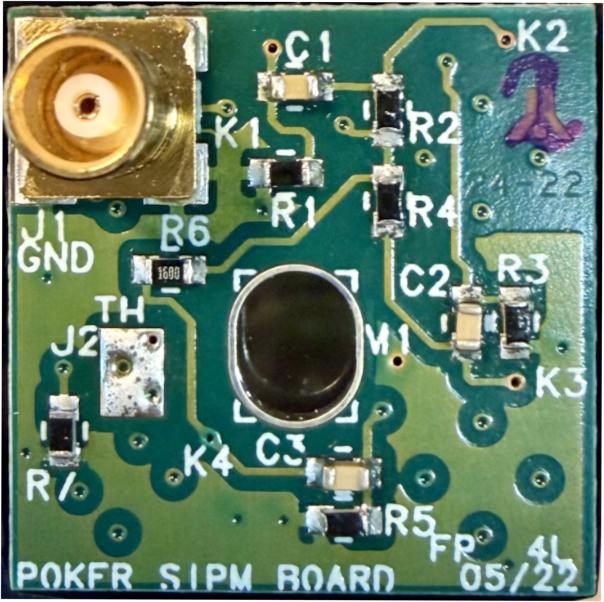}
    \caption{\textbf{Left:} picture of a \texttt{PKR-CAL-SiPM} printed circuit board, not yet populated with the photo-detectors. The passive components and the output connector are hosted on the opposite PCB face. The large rhomboidal pads are connected to the SiPMs thermal junctions, to extract heat from the devices during operation. \textbf{Right:} a fully-assembled \texttt{PKR-CAL-SiPM board}. SiPMs are hosted on the opposite face. The central hole is used to host a SMA connector to attach to the sensor an optical fibre from the laser monitoring system.}
    \label{fig:PKR-CAL-SiPM2}
\end{figure*}

A picture of the actual PCB circuit where the four SiPMs are hosted is shown in Fig.~\ref{fig:PKR-CAL-SiPM2}, where the soldering pads for the surface-mount SiPMs are clearly visible, together with the large thermal release pads. The passive components are mounted on the opposite layer. The dimension of the circuit (1.98 x 1.98 cm$^2$) are matched to the small face of the PbWO$_4$ crystal where the sensor is glued. The central hole in the PCB accommodates a custom connector used to couple an optical fiber from the PKR-CAL laser monitoring system. A Pt100 SMD sensor is mounted on the \texttt{PKR-CAL-SiPM} PCB close to the SiPMs to monitor their temperature.

\subsection{Laboratory characterization setup}\label{sec:labsetup}

\begin{figure*}[t]
\centering
{\includegraphics[width=.7\textwidth]{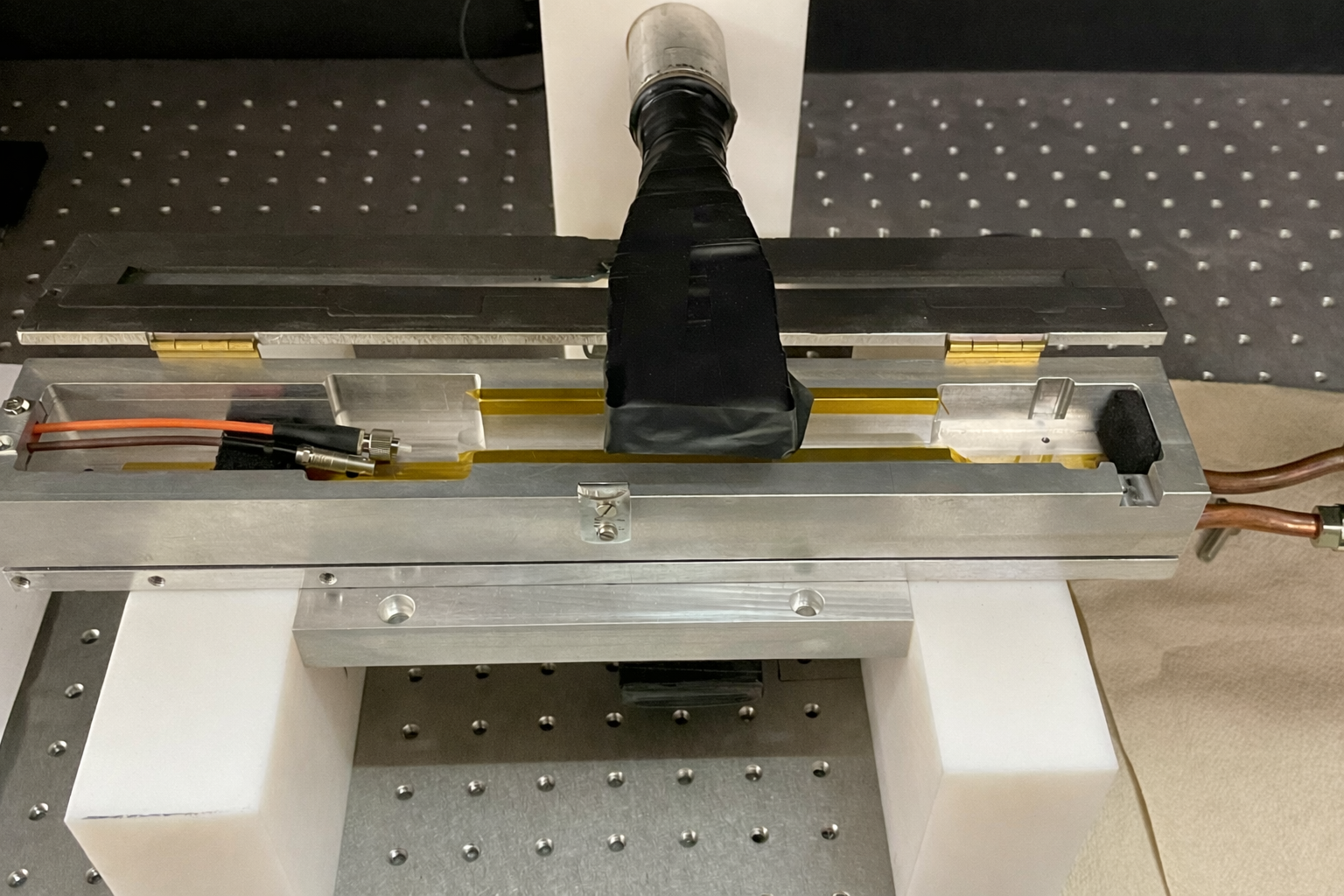}}
\caption{Picture of the setup employed for laboratory characterization of the \texttt{PKR-CAL-SiPM} device. A PbWO$_4$ crystal with a \texttt{PKR-CAL-SiPM} sensor was inserted in the aluminium frame, and the light pulses from a pulsed laser were transported to it via an optical fibre. The two plastic scintillator counters shown in the picture were employed for a different set of measurements exploiting cosmic-rays, not discussed in this work.\label{fig:setup}}
\end{figure*}

To characterize the main properties of the \texttt{PKR-CAL-SiPM} device, we implemented a dedicated setup based on a full PKR-CAL single-channel assembly, with the photosensor coupled to a PbWO$_4$ crystal. We measured the sensor response to light pulses generated by a very fast pulsed laser (Hamamatsu PLP-10, 440 nm) under different intensity and frequency conditions.  The typical duration of each laser pulse is approximately 70 ps. A picture of the setup is shown in Fig.~\ref{fig:setup}.  The light from the laser was transported with an optical fiber to a passive 50/50 splitter. One of the outputs was connected to a power
meter (Newport M 2930 C, with $\pm 2\%$ accuracy) to monitor the laser stability. The second splitter output was coupled to an optical fibre, with the other end inserted inside the hole of the \texttt{PKR-CAL-SiPM}.  The \texttt{PKR-CAL-SiPM} was attached to the PbWO$_4$ crystal using an optical gel, with a reflective VM2000 leaflet glued on the opposite face to maximize the reflection of the laser light. This setup ensured that the light spot from the laser, after reflecting on the opposite face the crystal, was uniformly illuminating the sensor surface. To ensure thermal stability, the crystal assembly was inserted in a aluminium enclosure, soldered to a cooling pipe connected to an external chiller.
The sensor was connected to a POKER amplifier, whose signal was acquired using a digital oscilloscope, triggered by the laser controller synchronization signal. For each data-taking run, the amplitude of the \texttt{PKR-CAL-SiPM} pulses was computed online via the oscilloscope, and the average value and RMS of the corresponding distributions were recorded. This setup was mostly used during the early R$\&$D phase to characterize individually the detector components. It was used to cross-check the high-frequency behaviour of the \texttt{PKR-CAL-SiPM}, discussed later in Seq.~\ref{sec:high_frec}.

\subsection{The EEE cosmic-ray test facility in Genova}

The POKERINO detector was commissioned in Genova, exploiting the existing cosmic ray test facility developed within the EEE project~\cite{LaRocca:2016xbr}. This facility is based on a single EEE telescope, made from three large-area (90$\times$180 cm$^{2}$) Multigap Resistive Plate Chambers (MRPCs) stacked vertically at a relative distance of 50 cm. The measurement of the impact point of a cosmic ray on each of the three chambers allows the reconstruction of the corresponding trajectory with spatial resolution of approximately 1 cm. 

Thanks to the large gap between the detector planes, a detector can be easily placed on top of any  EEE chamber. In this configuration, the reconstructed trajectories can be used to evaluate the hit point of the cosmic ray on the studied detector. The EEE chamber can thus be exploited as a ``detector characterization facility''~\cite{Grazzi:2023drv} by measuring and correlating the two detectors responses. When a coincidence signal among the three chambers is detected, the EEE DAQ system records data from the three MRPCs. In addition, the system provides the absolute event time at which the event occurs, thanks to the connection to an external GPS. This allows to perform the detector characterization independently from ongoing EEE operations, exploiting an external readout system, provided that the latter also provides a measurement of the absolute event time. For the POKERINO characterization, we employed a Waveboard-based readout, with the ADC operating in self-trigger mode. During the offline analysis, the two datasets were analysed simultaneously, searching for events in coincidence within a narrow time window. 

\subsection{Experimental setup at the CERN SPS H6 beamline}

The H6 beamline at CERN North Area ``Experimental Hall 1'' (ENH1) is a versatile beamline that provides user with hadron, muon, or electron beams in a broad momentum range (10-200 GeV/c), with intensity up to $10^7$ ($10^5$) hadrons (electrons) per SPS spill of 4.8 seconds~\cite{Banerjee:2774716}. The H6 beam is produced by having the primary 400 GeV/c proton beam from the SPS accelerator impinging on the thick beryllium target T4, and then filtering the momentum and particle species of produced secondary through a set of subsequent dipole magnets, collimators, and targets. In particular, a low-intensity, high-purity electron beam is obtained by selecting negative particles emerging from the target within a given momentum range through a set of magnets, and having these passing through a thin radiator. After this, thanks to Bremsstrahlung emission, electrons are separated in momentum from hadrons, and can thus be further selected. The obtained $e^-$ beam purity is typically greater than $90~\%$, with an intensity up to 10$^4$ particles/spill, depending on the overall setup~\cite{CERN_H6_manual}. The intrinsic momentum beam spread can be adjusted by appropriate momentum-defining collimators down to sub-percent values\footnote{An effective parameterization of the beam momentum spread full width (assuming a rectangular distribution) reads~\cite{CERN_H6_manual}:
\begin{equation}\label{eq:H6momentum}
\frac{\Delta p}{p} = \frac{\sqrt{C3^2+C8^2}}{19.4}\, \% \; ,
\end{equation}
where $C3$ and $C8$ are the full width opening of the two momentum defining collimators installed on the beamline, expressed in mm.}.

\begin{figure*}[t]
    \centering
    \includegraphics[width=.5\textwidth]{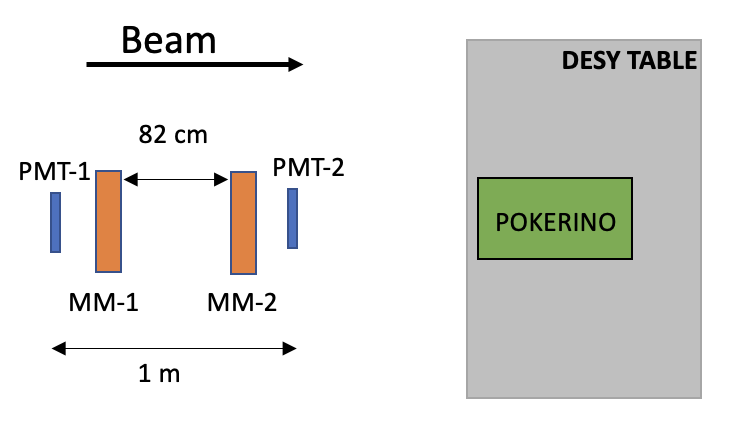}
    \quad
    \includegraphics[width=.4\textwidth]{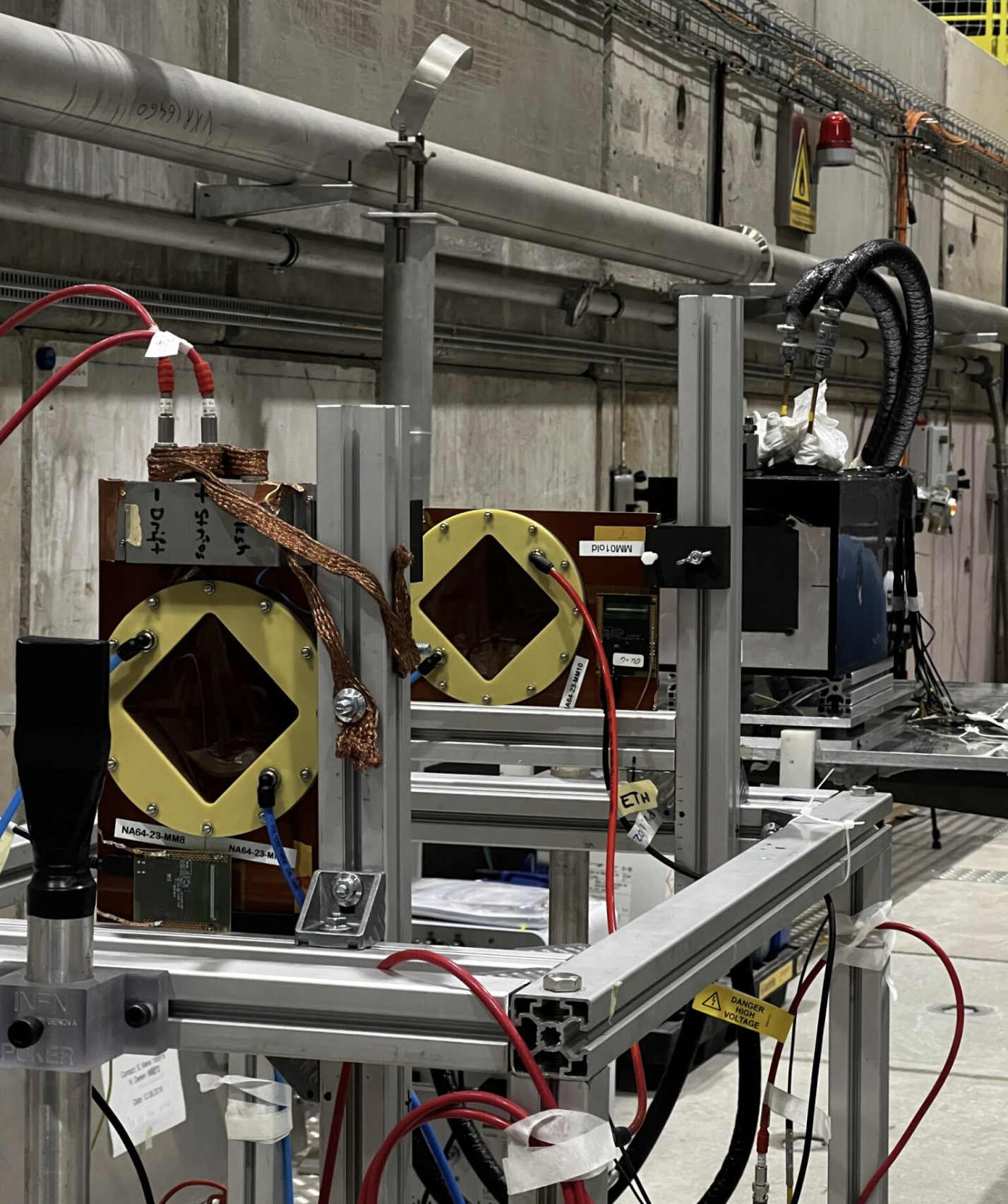}\\
    \caption{(\textbf{a}) Simplified scheme of the experimental setup used during the POKERINO 2024 test at CERN H6 beamline. \textbf{b}) Picture of the POKERINO setup during the 2024 test at CERN H6 beamline. From left to right, the first trigger scintillator, the two MM detectors, the second trigger scintillator, and the POKERINO detector placed on the DESY table.}
    \label{fig:POKERINO_setup_scheme_2024}
\end{figure*}

A dedicated POKERINO prototype characterization campaign was conducted in summer 2024 at the H6 beamline. A schematic of the setup is reported in Fig.~\ref{fig:POKERINO_setup_scheme_2024}, together with a picture of the experimental area with all detectors installed.  Two upstream PMT-based plastic scintillators counters (PMT-1 and PMT-2), $5\times5\times1.5$ cm$^3$ each were installed upstream along the beamline, at a relative distance of about one meter, to detect the impinging particles. Two MicroMegas (MM) detectors were installed in between the counters to measure the impinging particle trajectory and the corresponding impact point on the calorimeter front-face.

The prototype was installed on a movable ``DESY'' table, allowing us to translate it in the plane orthogonal to the beam (X-Y plane in the figure), thus changing the particles impact point on the detector. On the table, POKERINO was mounted on a rotating platform to rotate it in a plane parallel to the ground (Y-Z plane in the setup scheme) and align it with the beam direction. During the whole measurement, the cooling system was set to keep the calorimeter's temperature at 20$^\circ$ C. To control and monitor all detector parameters such as the \texttt{PKR-CAL-SiPM} bias voltages and currents, temperatures, DAQ status and rates, \ldots, the setup also included a preliminary version of the PKR-CAL EPICS-based controls system~\cite{Dalesio:1994qp}.

Signals from POKERINO and from the two upstream PMTs were connected to the experiment DAQ system. The MM readout was based on the APV-25 ASIC~\cite{French:2001xb}, handled by an independent DAQ system based on the CERN SRS system~\cite{Scharenberg:2022hyf}. The coincidence of the two beam-defining plastic scintillator counters provided a common trigger signal for both acquisition systems. For each run, events from the two datasets were matched during the offline analysis based on their time relative to the run start.

\section{Results}

In this section, we first discuss the POKERINO commissioning performed by measuring the prototype response to cosmic rays through the EEE-based facility installed in Genova, and we then present the results obtained from the characterization of the POKERINO prototype with high energy electrons and hadrons at CERN H6 beamline. These measurements represent a follow-up of the studies performed in 2023 with a preliminary version of the prototype at the H8 beamline~\cite{Antonov:2024glv}. During all test-beam measurements, the \texttt{PKR-CAL-SiPM} sensors were operated at their nominal working point $V_b=V_{BD}+5~$V, and the detector temperature was set at $20^\circ$C via the external chiller. The temperature of the detector was continuously monitored through a CYD218 module interfaced to the Pt100 sensors mounted within the detector, and observed to be stable within 0.2~$^\circ$C. The typical beam intensity was of the order of $10^4$ particles/spill.

\subsection{Cosmic-rays commissioning}

During the commissioning phase, the POKERINO detector was installed within the EEE-based cosmic ray-telescope, on top of the uppermost chamber. The PbWO$_4$ crystals were aligned with their long axis along the EEE Y axis, with the \texttt{PKR-CAL-SiPM} sensor close to the low-Y edge. Different data-taking runs were collected, operating the detector with different bias voltages, to study each channel's response. On average, each run lasted for few days. During all tests, the chiller water temperature was set to 20 $^\circ$C.

\begin{figure*}[t]
    \centering
    \includegraphics[width=.45\textwidth]{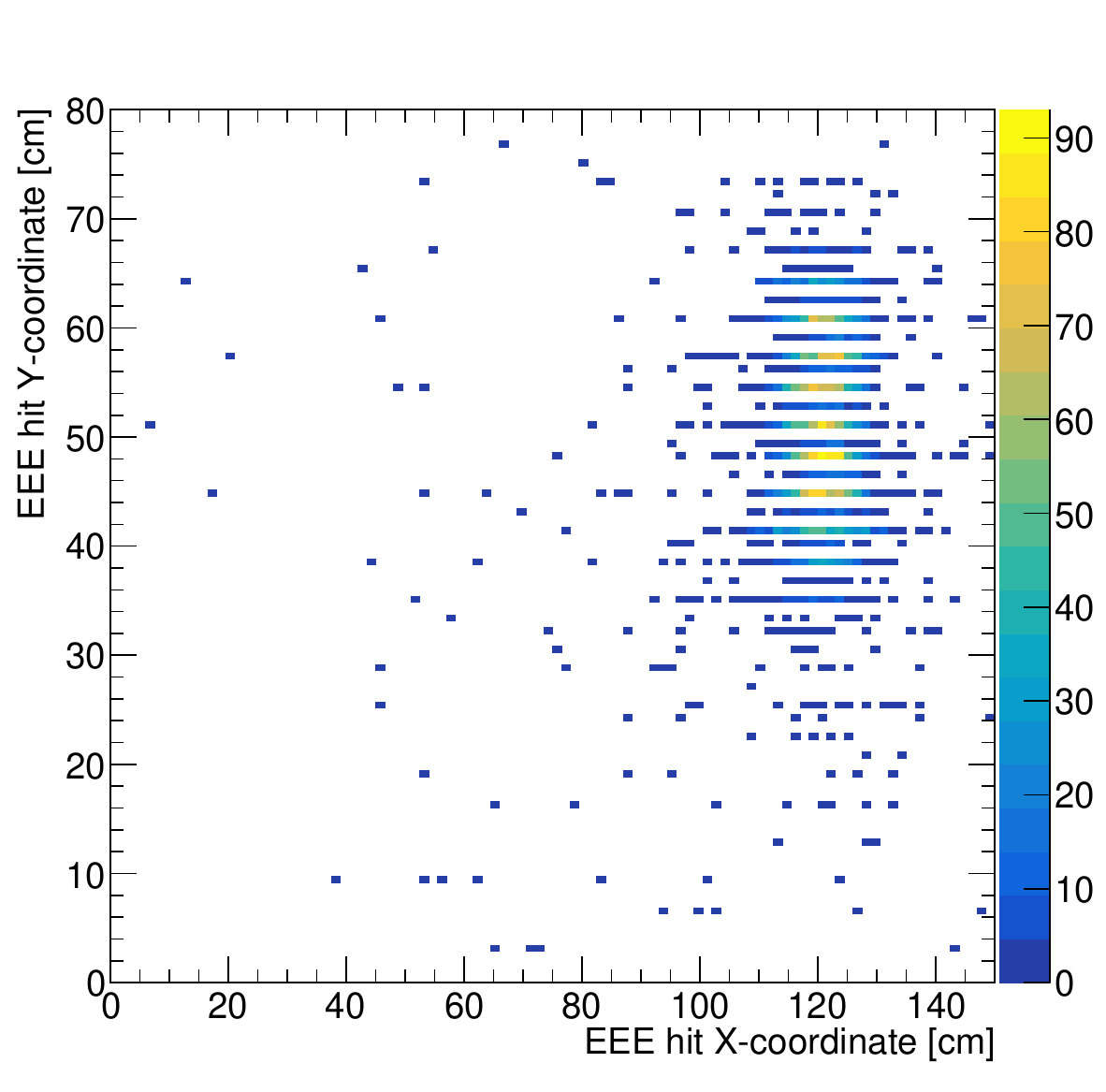}
    \quad
    \includegraphics[width=.45\textwidth]{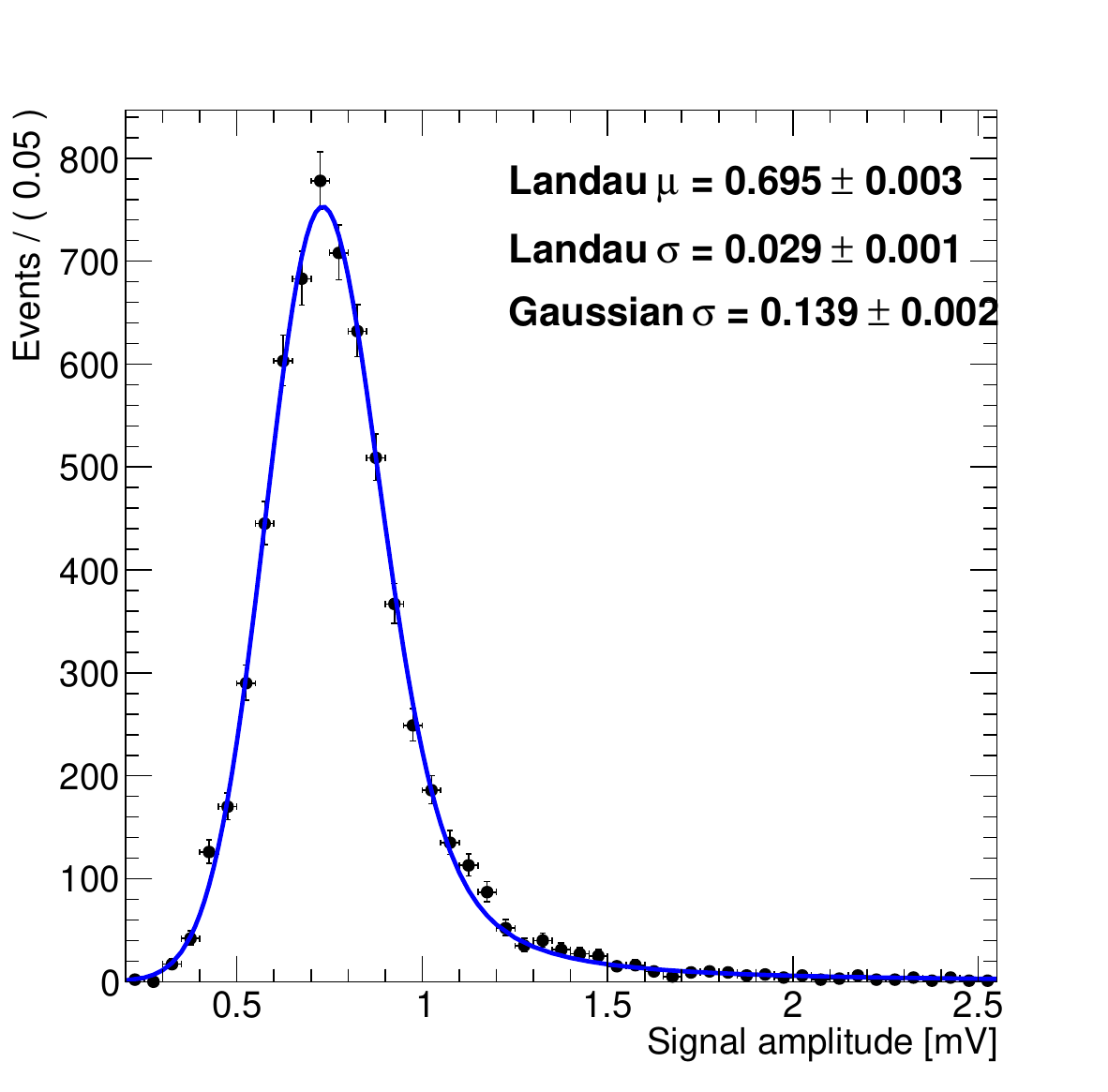}\\
    \caption{(\textbf{a}) XY coordinates of the hits measured by the top EEE chamber for cosmic-ray events selected through the cuts described in the text, with the further requirement of at least one signal from POKERINO crystals (see text for further details). \textbf{b}) Signal amplitude distribution of selected cosmic-ray events for the POKERINO centermost crystal.}
    \label{fig:ProtoCosmic}
\end{figure*}

Recorded data was processed offline to select a clean set of events with a vertical cosmic-ray muon passing through the POKERINO active volume. The selection criteria required, for each EEE event, the presence of a single down-going cosmic-ray track, with vertical direction cosine larger than 0.95, in time coincidence with a single event recorded by the POKERINO DAQ system. Figure~\ref{fig:ProtoCosmic}, left panel, shows the EEE top chamber XY hit position for events with at least one signal from POKERINO crystals - the shape of the latter is visible. Finally, for each crystal in a given column, events with the reconstructed EEE top chamber XY hit position matched to the column position were selected, and the corresponding amplitude distribution calculated. All distributions showed a clear peak from MIP-like energy deposition from cosmic rays -  for illustration, Fig.~\ref{fig:ProtoCosmic}, right panel, reports result obtained for the center-most crystal for bias voltage $V_{b}=V_{BD}+10.17$ V. To extract the peak position, a maximum likelihood unbinned fit was performed with a Landau function convoluted with a Gaussian resolution model. The most probable value of the Landau distribution was found to be approximately of about 0.65 mV for all channels, with an overall variation of about $10$\%. Monte Carlo simulations predicted a most probable energy deposition value for cosmic rays of about 20 MeV, resulting to an overall response of about 32.5 mV/GeV for this bias voltage value. This result, combined with other ancillary measurements not discussed in this document, returned an estimated light yield of about 5 phe/MeV. With these tests, we confirmed the proper operation of all prototype channels, obtaining a preliminary energy calibration value.

\subsection{Energy calibration}

During the test beam, the response of each POKERINO channel was calibrated exploiting data collected during a set of low-energy runs, in which the position of the detector was changed through the DESY platform to have the beam impinging on the centre of a given $(x,y)$ cell. For each channel, the energy calibration constant was determined by measuring the detector response to a 120 GeV/c $\mu^-$ beam, by centring the front face of each crystal on the beam spot. From Monte Carlo simulations, we estimated the most probable value of the energy deposition to be 260 MeV. In this experimental condition, the SiPMs sensors saturation effect was thus negligible: assuming, conservatively, a light yield of about 5~phe/MeV, the expected number of active \texttt{PKR-CAL-SiPM} cells was approximately 1300, resulting to an average occupancy of about 0.09$\%$.

\begin{figure*}
    \centering
    \includegraphics[width=0.7\textwidth]{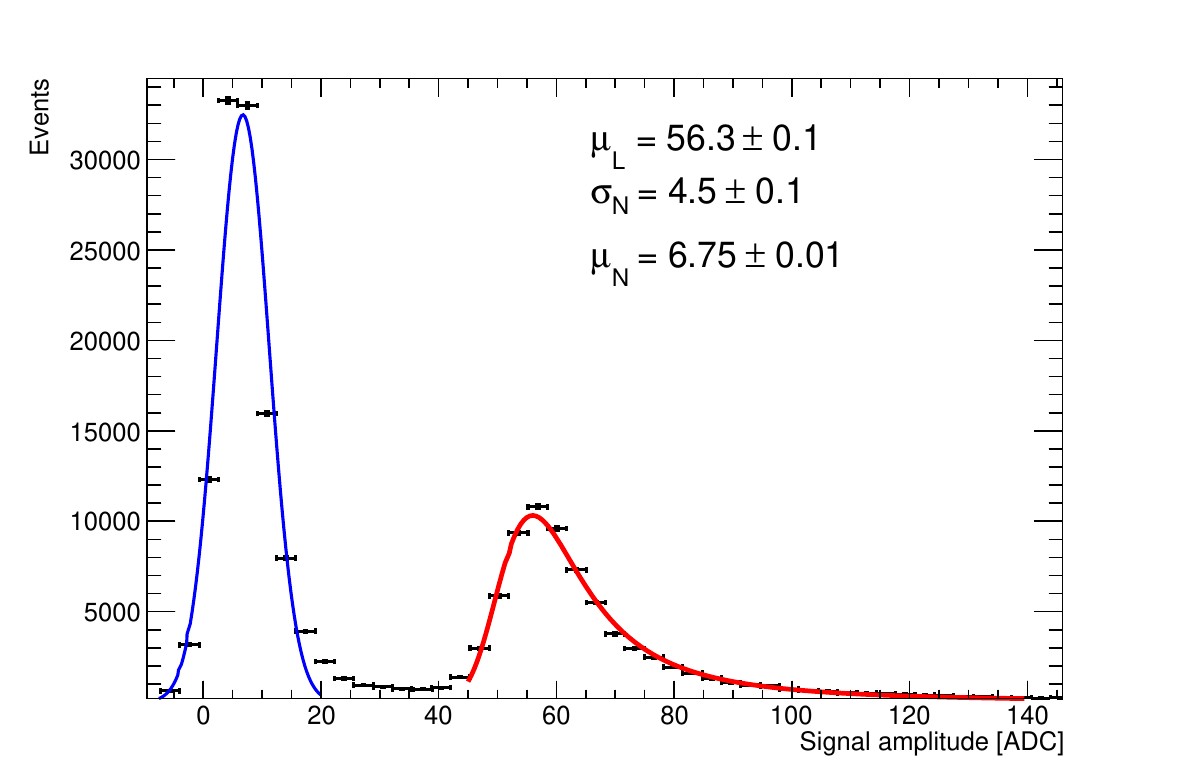}
    \caption{Energy deposition in the POKERINO central cell from 120 GeV/c $\mu^-$ impinging on it, in raw ADC units (after pedestal subtraction). See text for further details.}
    \label{fig:protoMuon}
\end{figure*}

Figure~\ref{fig:protoMuon} shows the signal amplitude distribution for the prototype central cell, in raw ADC units. A clear peak corresponding to the ionization by muons is visible, together with a low-energy peak due to noise.  Each data distribution was fitted with a Landau function convoluted with a Gaussian to extract the most probable value and thus the energy calibration constant $c$. The low-energy peak was fitted with a Gaussian function to measure the equivalent noise energy, conservatively estimated as $ENE=(\mu_N+3\sigma_N)\times c$, where $\mu_N$ and $\sigma_N$ are, respectively, the mean value and the standard deviation of the fit function. For all cells, we found $ENE\approx100$~MeV.

\subsection{Linearity study and energy correction}

To study the effect of the finite number of active cells in \texttt{PKR-CAL-SiPM} photosensors and derive an appropriate correction to saturation effects, we measured the linearity of the detector by collecting various acquisition runs at different electron-beam energy, ranging from 10~GeV to 100~GeV at 10~GeV steps, and comparing the results with the predictions from Monte Carlo. In these runs, the beam impinged on the centre of the POKERINO central cell. The observed purity of the electron beam, estimated from the measured spectrum from the relative yield of full-energy deposition events was more than 90$\%$ for all momentum values.

To minimize the intrinsic momentum spread of the beam, correlated with the beam spatial dimensions, we optimized the H6 configuration by closing the two beam-defining collimators from their nominal $\pm 20$~mm setting to the value of $\pm 5$~mm. To identify a clean set of events for the comparison, avoiding biases due to an inaccurate description of the beam shape in Monte Carlo, we exploited the hit position information provided by the two MM detectors, extrapolating the straight trajectory to the calorimeter front face and introducing a 3-$mm$ cut on the distance between the extrapolated hit position and the seed cell centre. The same procedure was implemented to the Monte Carlo dataset, in which, for simplicity, we employed the generator-level extrapolated track point, without including the MM detectors resolution effects.

\begin{figure*}[t]
    \centering
    \includegraphics[width=.47\textwidth]{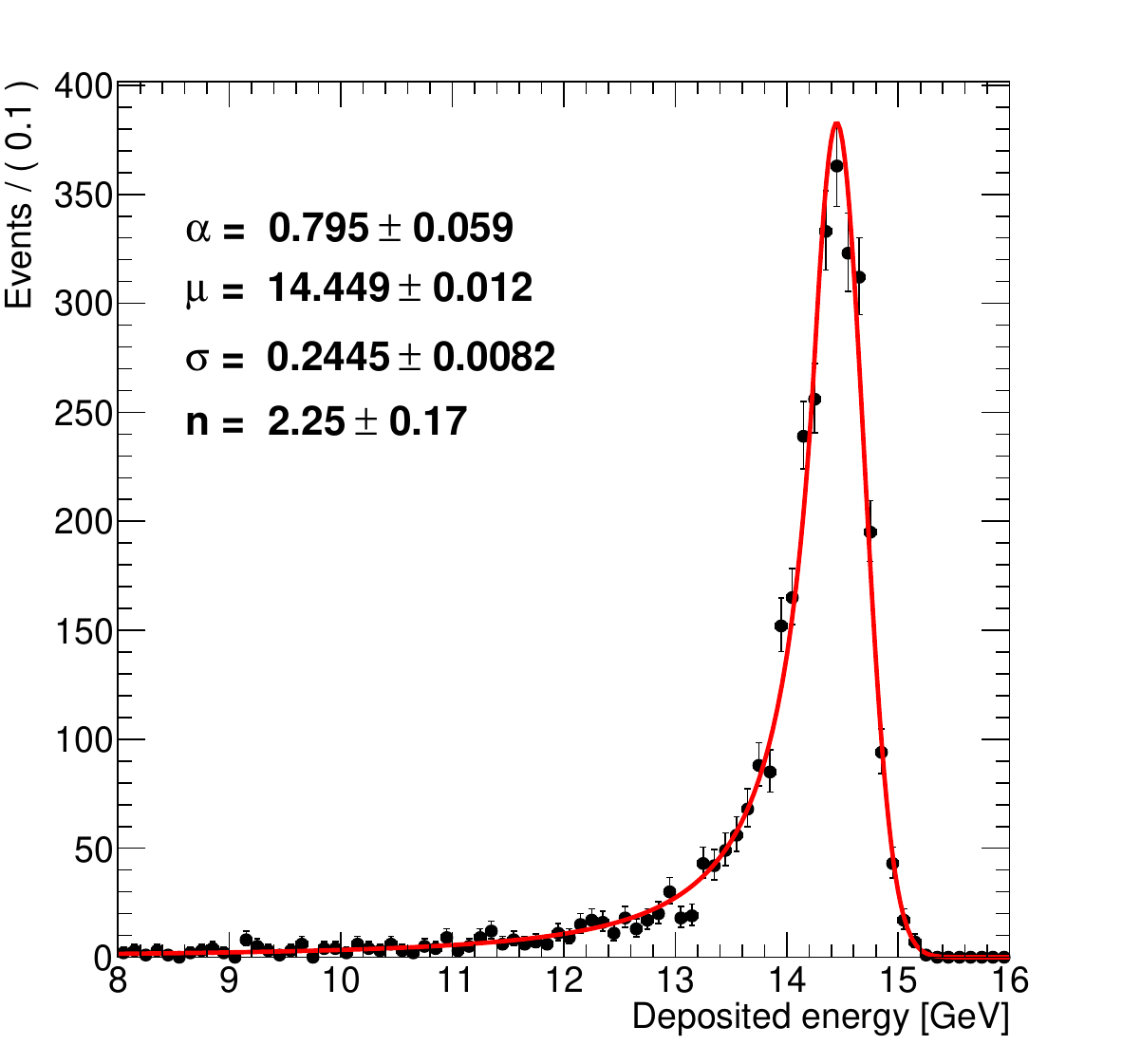}
    \quad
    \includegraphics[width=.47\textwidth]{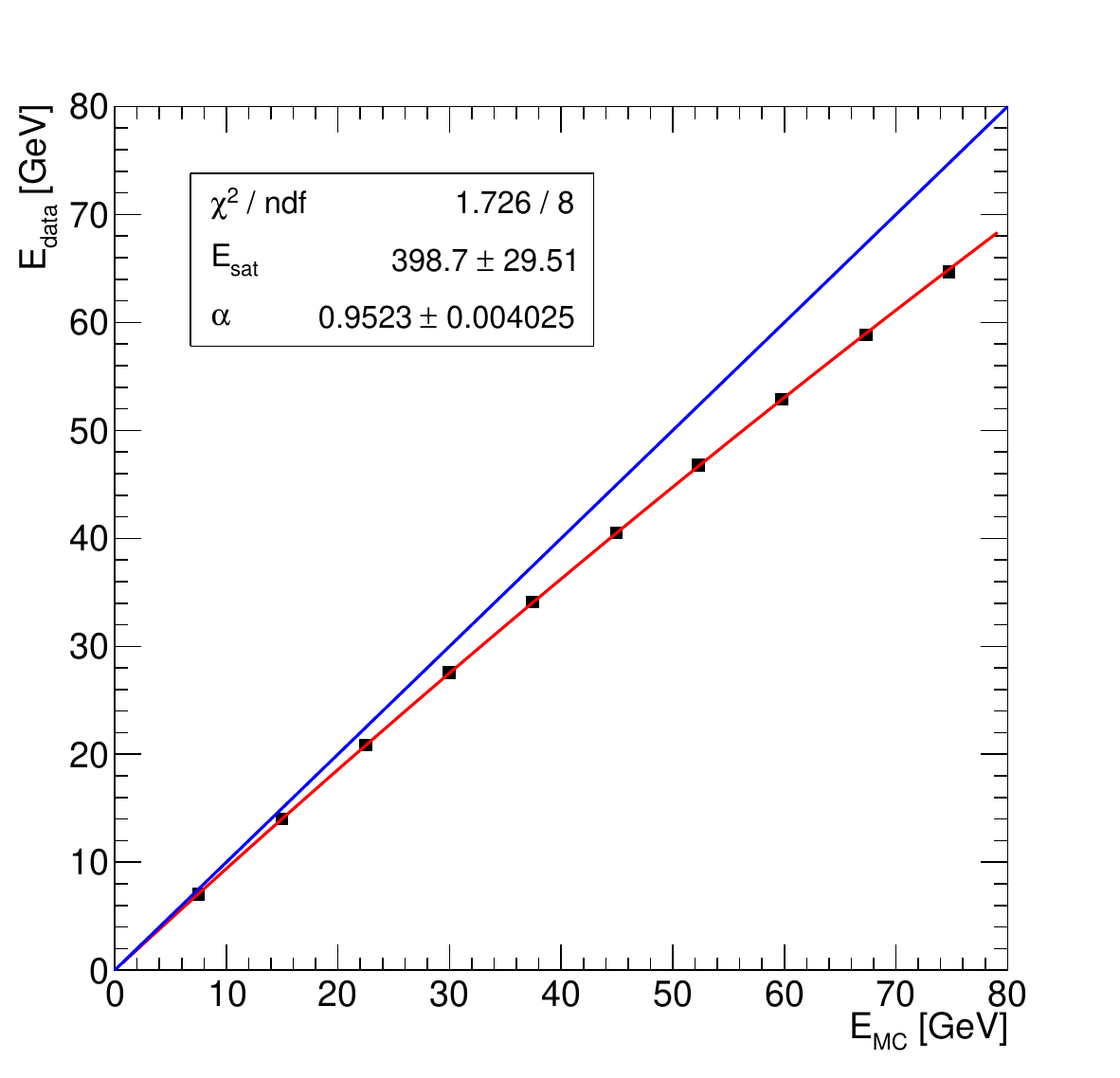}\\
    \caption{(\textbf{a}) Energy distribution for POKERINO central cell hit by a 20 GeV/c $e^+$ beam. \textbf{b}) Energy deposition in POKERINO central cell for different impinging $e^+$ beam energy, as a function of the ``true'' energy deposition predicted by Monte Carlo. The red line is the result of a $\chi^2$ fit performed with the function $E_{data}=\alpha E_{sat}(1-\exp{(-E_{true}/E_{sat}))}$ where $E_{true}$ is the true energy deposited if no saturation effects were present. The blue dashed line to guide the eye corresponds to the ideal case $E_{data}=E_{true}$.}
    \label{fig:POKERINO_E_2024}
\end{figure*}

For each run, we measured the energy deposition in the POKERINO central cell, and we performed a fit to the corresponding distribution with a \texttt{RooFit RooCrystalBall} PDF. As an example, Fig.~\ref{fig:POKERINO_E_2024}, left panel, shows the obtained result for the 20-GeV run. To assess the systematic uncertainty of the measurements, two independent runs at 10 GeV/c were collected at different times. The corresponding seed energy mean values, $\mu_1=(7.074\pm0.013)$~GeV and $\mu_2=(7.013\pm0.014)$~GeV, were found to be different by more than $3\sigma$. A systematic uncertainty of $0.6\%$ was therefore assigned to each measured mean value, corresponding to the relative value of the observed difference, subtracted by the statistical uncertainty contribution. This estimate is supported by an independent analysis of the spill-by-spill fluctuations of $\mu$, whose standard deviation was found to be of the same order of magnitude.
We scrutinized the possible origin of this effect by investigating whether it could be attributed to variations in the beam impact position or beam-spot properties between the two runs. Although the 3-mm cut on the reconstructed impact point was applied in the analysis to mitigate such effects, we performed an additional cross-check by comparing the total energy deposited in POKERINO for the two measurements, since this observable is less sensitive to small variations in the beam position within the matrix. The corresponding total energy mean values were found to be compatible within their uncertainties, indicating that the observed difference in the seed energy is not associated with a change in the overall calorimeter response, but is most likely due to small differences in the beam conditions. 

The obtained mean energy values were compared from the predictions from Monte Carlo simulations, in which no saturation effects are present. Figure~\ref{fig:POKERINO_E_2024}, right panel, reports the result for the POKERINO central cell, for which the largest energy interval was scrutinized. The data points show a deviation from linearity, with a clear saturation trend. We parameterized this through an exponential function 
\begin{equation}\label{eq:fit}
    E=\alpha \, E_{sat}\times(1-\exp(-E_{true}/E_{sat})) \; \;,
\end{equation}
with $\alpha$ and $E_{sat}$ free parameters. Here, $E_{sat}$ parametrizes the maximum equivalent energy that can be recorded by the \texttt{PKR-CAL-SiPM} due to SiPM cells saturation, while $\alpha$ is a correction term to the energy calibration coefficient. The result of a $\chi^2$ fit performed with this function to the data is reported on the same figure. The obtained value for $E_{sat}=(395\pm15)$~GeV can be related to the detector properties observing that the average number $N^\prime$ of activated SiPM cells due a light pulse with $N_\gamma$ optical photons is given by the formula $N^\prime=N_{c}(1-\exp(-\mu/N_c))$, where $N_c$ is the total number of available cells and $\mu \equiv N_\gamma \times \varepsilon$, $\varepsilon$ being the SiPM photon detection efficiency. The function used for the fit (Eq.~\ref{eq:fit}) corresponds to this model by recognizing $E_{sat}\equiv N_c/LY$, where $LY$ is the calorimeter cell overall light yield, being $E= N^\prime / LY$. From the $E_{sat}$ value obtained from the fit, and considering the light yield of about 5 phe/MeV, measured during the detector commissioning in Genova, we compute the effective number of total cells $N_C\approx 2\times10^6$. This number is  $40\%$ higher than the real number of cells in the \texttt{PKR-CAL-SiPM} sensor, as predicted by a more refined saturation model in which cells re-charge and re-triggering effects are included~\cite{Rosado:2019osg}.

The same study was repeated for the 8 periphery cell - due to the limited run time, their response was characterized only for lower beam energies, at 10~GeV, 20~GeV, and 40~GeV -- this was motivated by the fact that, during the measurements with the beam impinging on the central cell, representative of the final POKER measurement setup, the energy released in the periphery cells was significantly smaller. Also in this case the two-parameters exponential function previously discussed was adopted to describe the saturation trend. We observed that all the saturation parameters $E_{sat}$ are compatible within their uncertainty, confirming the uniformity of the individual cell properties.

\subsection{Energy resolution}

\begin{figure*}[t]
    \centering
    \includegraphics[width=0.8\textwidth]{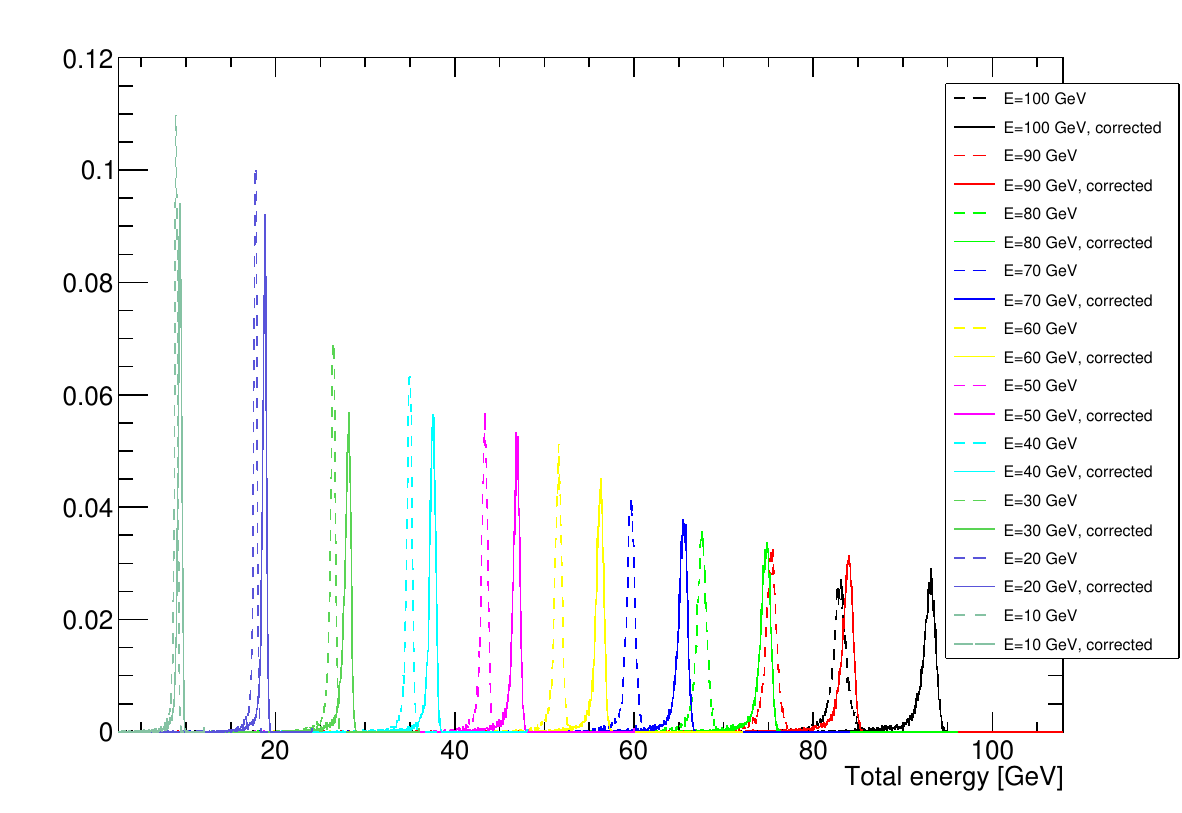}
    \caption{The energy spectra measured by POKERINO exploiting $e^+$ beams at different energies, as shown in the legend, including (continuous line) or not (dashed line) the energy saturation correction. See text for further details.}
    \label{POKERINO_energy_2024}
\end{figure*}

The POKERINO energy resolution $\sigma_E/E$ was characterized using the same dataset employed for the linearity studies, with the $e^+$ beam impinging on the central cell. For each cell, an event-by-event energy correction was applied to account for saturation effects, introducing a new variable $E^\prime_i=-{E_{sat-i}} \ln(1-E_i/(\alpha_i E_{sat-i})$, where $E_{sat-i}$ is the energy saturation parameter of the cell determined through the procedure discussed before. To reproduce the high-intensity conditions expected in POKER, in which the expected beam-spot radius is of the order of one cm, no selection cuts on the beam impact point on the detector was included. 
Figure~\ref{POKERINO_energy_2024} shows the obtained results, comparing for each beam setting the total energy distribution in POKERINO including (full lines) or not (dashed lines) the saturation correction. 

\begin{figure*}[t]
    \centering
    \includegraphics[width=0.8\textwidth]{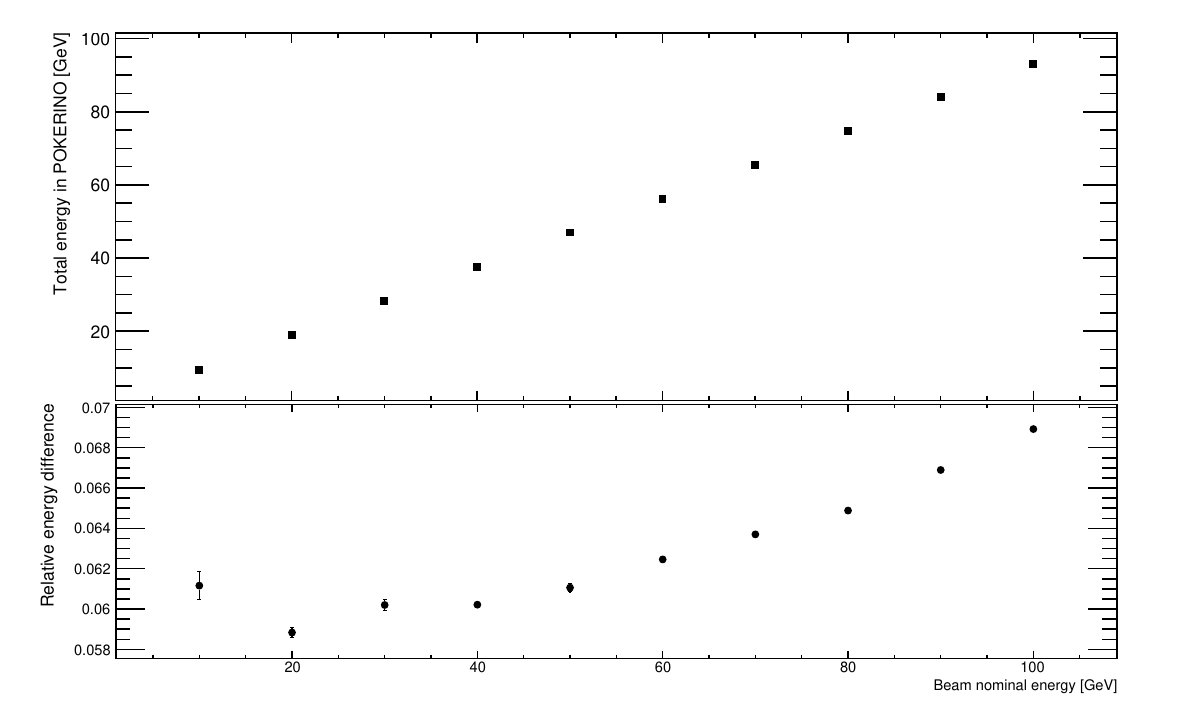}
    \caption{Top plot: total energy measured in POKERINO as a function of the nominal beam energy. Bottom plot: relative difference between the nominal beam energy and that measured in POKERINO as a function of the nominal beam energy.}
    \label{fig:plt_POKER_line_2024}
\end{figure*}

\begin{figure*}[t]
    \centering
    \includegraphics[width=0.8\textwidth]{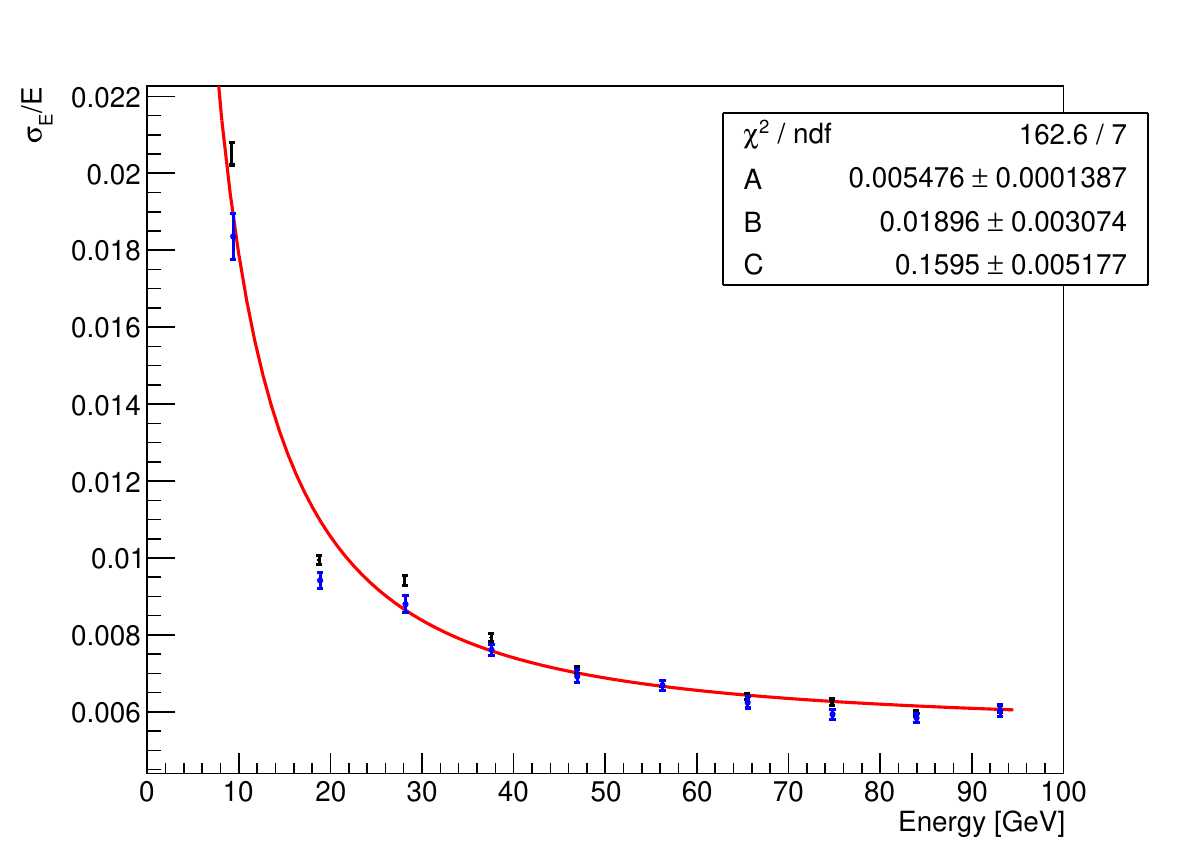}
    \caption{POKERINO energy resolution $\sigma/E$ as a function of the total energy deposition, after introducing an event-by-event correction to each cell measured energy to account for SiPM saturation effects. The blue (black) points correspond to the results obtained with (without) a cut on the particle impact point on the detector. The red curve is the result of the best fit to data-points not including the impact-point cut with the resolution model discussed in the text.}
    \label{fig:plt_POKER_reso_2024}
\end{figure*}

Each energy-corrected distribution was fit through a \texttt{Roofit RooCrystalBall} PDF to determine the average energy deposition and the corresponding resolution. The linearity of the POKERINO response is shown in Fig.~\ref{fig:plt_POKER_line_2024}, in which the top (bottom) plot shows the correlation between the average energy deposition (relative difference between the nominal beam energy and the average energy deposition) and the nominal beam energy, respectively. Since no energy-leakage corrections are applied, the relative energy difference is of about 0.065, with a $\approx 10\%$ variation in the considered energy interval -- we anticipate to correct this effect in POKER through an ad-hoc energy correction mechanism derived from MC simulations.
The relative energy resolution is shown in Fig.~\ref{fig:plt_POKER_reso_2024}, reporting the $\sigma/E$ observable as a function of the total measured energy.  The data was parameterized via a function $\sigma/E=A \oplus B/\sqrt{E} \oplus C/E$, where $\oplus$ denotes a quadratic sum. The obtained value of the systematic parameter $A$ was $(0.55 \pm 0.01)\%$, compatible with the design requirements. Similarly, the statistical term $B$ was $(1.8 \pm 0.5) \%$, confirming the overall light collection yield of about $3\div 4$ phe/MeV. The noise term $C=(161\pm5)$~MeV is also compatible with the conservative single-cell ENE estimate previously discussed. The measured point at 10 GeV beam energy deviates from the overall trend visible at higher energy; we observe that this energy value corresponds to the minimum momentum acceptance of the H6 beamline, and thus possible spurious effects on the momentum spread or on the nominal value are possible. If this point is excluded from the fit, the result shows a lower $\chi^2/NDF=7.8$ value, with the constant resolution term reducing to $A=(0.41\pm0.01)\%$, at the price of a slightly larger value for the statistical term $B$.

All measurements discussed previously were performed closing the two H6 momentum-defining collimators to $\pm$5 mm - in this configuration, the expected momentum spread width (assuming a rectangular PDF) is 0.7$\%$, with an equivalent standard deviation $\sigma_p/p=0.2\%$. To check the possible effect of the momentum spread on the energy resolution, we repeated the analysis by selecting events with projected $e^+$ impact point within 3 mm from the crystal centre (see again Fig.~\ref{fig:plt_POKER_reso_2024}, blue points). In general, the two datasets agree quite well, a part from at low energy, where the point from the filtered dataset is characterized by a lower relative resolution - this supports our explanation of the resolution behaviour at low energy. If the lowest energy point is again excluded from the analysis, the fit result yields $A=(0.43\pm0.02)\%$ and $B=(3.8\pm0.1)\%$, with $\chi^2/NDF=2.3$. The value of the $A$ parameter is compatible with the result obtained previously, supporting the fact that, in both scenarios, the intrinsic beam spread is smaller than the calorimeter resolution.

\subsection{\label{sec:high_frec}Response to high-frequency beams}

In POKER, the \texttt{PKR-CAL-SiPM} sensors are exposed to the intense scintillation light from PbWO$_4$ crystals induced by impinging 100 GeV/c positrons at average rate in range 100 kHz - 1 MHz, resulting in a sizeable DC current flowing across the bias resistors. Due to the voltage drop on the latter, the effective bias voltage of the photo-sensors, and hence the gain, is reduced. This effect leads to a dependency of the SiPM response on the beam average intensity $f_b$, and, most important, in case of any sudden variation of the impinging particle rate may introduce gain fluctuations affecting the overall detector resolution. 

The magnitude of this effect can be estimated through a simplified DC model of the \texttt{PRK-CAL-SiPM} devices in which the intrinsic particle detection efficiency dependency on the bias voltage is neglected. Qualitatively, the process is governed by a negative-feedback mechanism: if the beam intensity suddenly increases, so does the bias current flowing through the sensor, and thus the voltage drop across the bias resistor grows. As consequence, the voltage difference across the sensor diminishes and the gain drops, resulting in a decrease of the bias current. Quantitively, at fixed temperature, the \texttt{PRK-CAL-SiPM} gain can be expressed as:
\begin{equation}
    G=g_0 (V_{b}-V_{BD}-I\,R_b) \; \;,
\end{equation}
where $V_{b}$ is the bias voltage provided by the generator, $V_{BD}$ is the SiPM breakdown voltage, $I$ is the average current flowing through the bias resistor, and $g_0$ is a proportionality constant, depending on the SiPM cells capacitance\footnote{For the S14160-6010 device, a fit to the data reported on the datasheet provides $g_0 \simeq 35000/V$.}. Here, $R_b$ is the overall detector bias resistance -- in the present setup, $R_b=60~\Omega$ is the parallel combination of the four 200 $\Omega$ bias resistors mounted on the \texttt{PKR-CAL-SiPM}, plus a 10~$\Omega$ contribution from the external bias circuit. Considering the average impinging particles rate $f_b$ and calling $N^\prime$ the average number of cells activated by the scintillation light pulses, the relation $I=G\cdot(e\,f N^\prime)$ holds, resulting to the following result:
\begin{align}
    G&=g_0 (V_{b}-V_{BD}-G \, e f_b N^\prime R_b) \nonumber \\ 
    G&=\frac{g_0(V_{b}-V_{BD})}{1+g_0 e f_b N^\prime R_b}\; \; .
\end{align}
This equation shows that, for beam rate $f_b \ll f_{cut}$, with $f_{cut}=(g_0 \, e N^\prime R_b)^{-1}$, a linear relation between the gain and $V_b$ holds, while at larger frequencies the gain drops as $f_{cut}/f_{b}$. For the PKR-CAL centre-most crystal, a conservative estimate for $N^\prime$ is $300\times 10^{3}$ phe, therefore $f_{cut}\approx 10 \,$MHz. Any instantaneous beam rate fluctuation $\sigma _{f_b}$ would induce a variation in the gain, and thus in the reconstructed energy response of the calorimeter, contributing to the overall energy resolution through a constant term
\begin{equation}
    \frac{\sigma_E}{E}=\frac{\sigma_G}{G}=\frac{\sigma_{f_b}}{f_b}\cdot\frac{1}{1+f_{cut}/f_b}\;\;.
\end{equation}
In the POKER experiment, the nominal value of $f_b$ is about 100 kHz, so that $f_{cut}/f_{b} \gtrsim 100$. Since at H4, for these operating conditions, the in-spill beam intensity fluctuations can reach up to $\sigma_{f_b}/f_b\approx 30\%$, this ensures that intensity-induced gain fluctuations contribute as a sub-dominant term to the overall energy resolution\footnote{We observe that, in principle, larger spill-by-spill variations of $f_b$ are possible at H4 - in POKER, we plan to compensate for this effect by recalibrating the central cell response for each spill, exploiting the signature of 100 GeV/c $e^+$ events.}. 

\begin{figure*}[t]
    \centering
    \includegraphics[width=.47\textwidth]{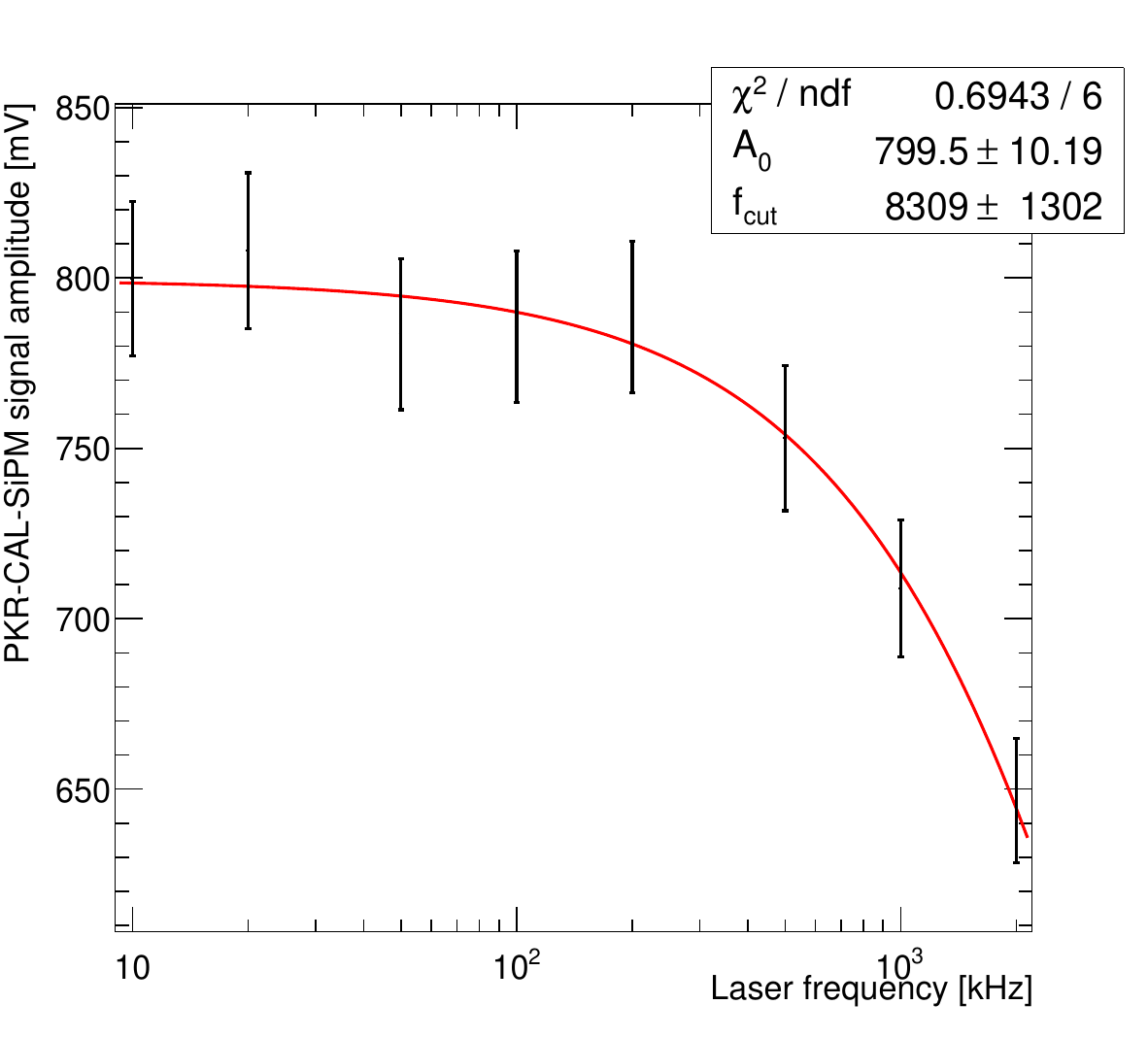}
    \quad
    \includegraphics[width=.47\textwidth]{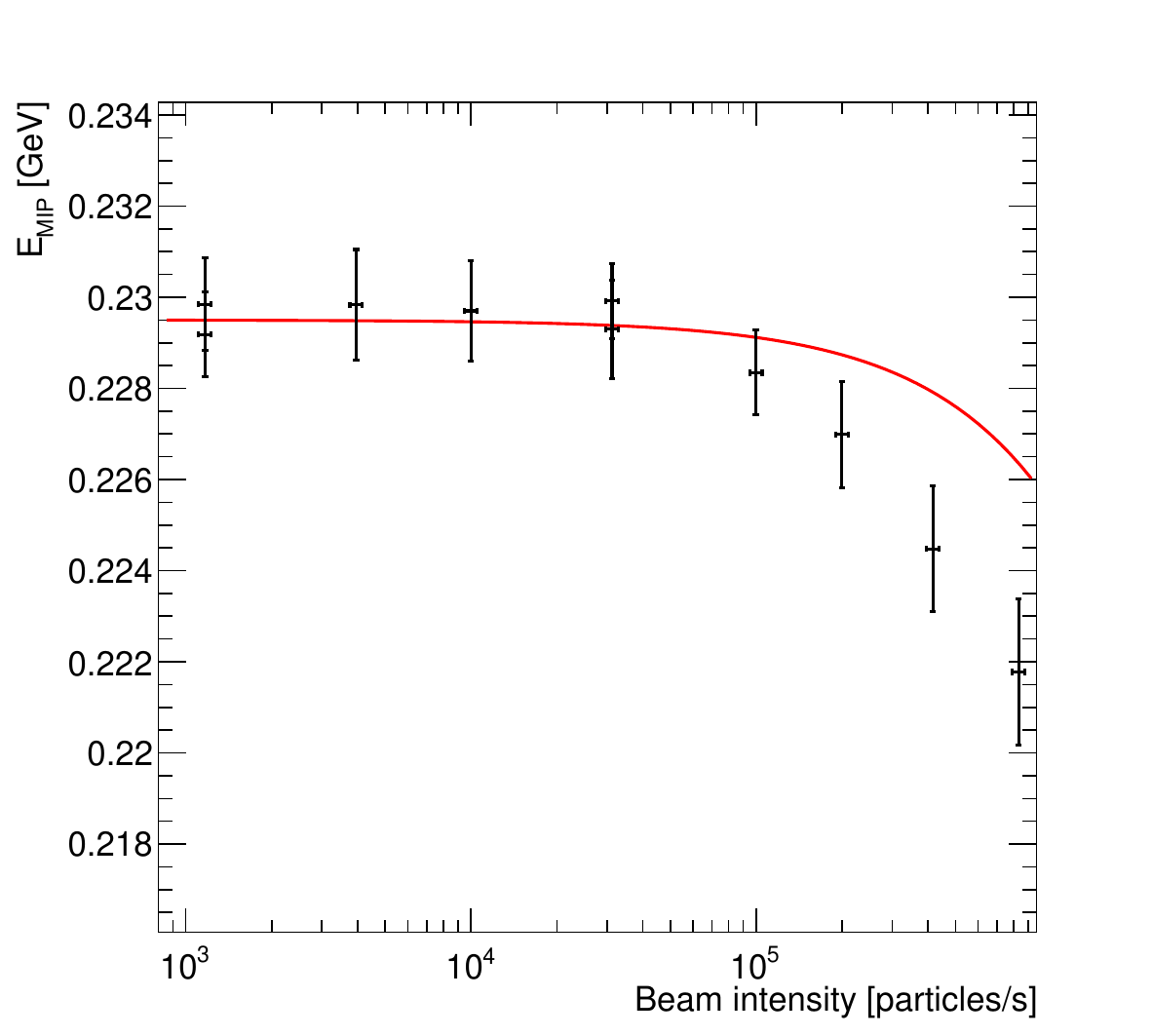}\\
    \caption{(\textbf{a}) Measured amplitude dependency of a \texttt{PKR-CAL-SiPM} sensor exposed to intense laser light pulses ($N^\prime \simeq 400\times 10^3$ phe), as a function of the laser frequency. (\textbf{b}) Value of the MIP Landau peak $E_{MIP}$ in the POKERINO central cell as a function of the beam intensity. The vertical bar errors correspond to the spill-by-spill standard deviation of $E_{MIP}$. See text for further details.}
    \label{fig:freq}
\end{figure*}

The validity of this model was checked with a dedicated measurement using the setup discussed in Sec.~\ref{sec:labsetup}. The pulsed laser intensity was set to obtain a \texttt{PKR-CAL-SiPM} signal equivalent to approximately $400\times10^{3}$ phe at low frequency, and the sensor response was then measured as a function of the laser pulses frequency $f_b$. In this configuration, the expected cut-off frequency is approximately $f_{cut}=8.9$ MHz. To correct any possible dependency of the laser itself on the latter observable, for each frequency value we measured the laser average power $P$ with the photodiode, and we applied a correction to account for any frequency-dependent variation of the light intensity through the relation $A=A_{raw} \times \frac{P_0/f_0}{P/f}$, where $P_0$ is the measured power at the lowest measured frequency $f_0=1$~kHz. The magnitude of this correction was frequency-dependent, with a maximum value of $9\%$ for $f=2$~MHz. The uncertainty on this correction factor, properly accounted for in the analysis, is dominated by the $2\%$ accuracy on the absolute laser power $P$ measured with the photodiode. The obtained result is show in Fig.~\ref{fig:freq}, left panel, reporting the experimental measurements together with a fit performed with the simplified-model function $A(f)=A_0/(1+f/f_{cut})$, with the two free parameters $A_0$ and $f_{cut}$. The measured cut-off frequency $f_{cut}=8.3\pm1.3$ MHz is in  good agreement with the model prediction, accounting for the uncertainty on $N^\prime$.

In order to verify the obtained results in a condition representative of the PKR-CAL foreseen measurement at H4, we performed a high-intensity test during the POKERINO-test beam, exploiting the H6 beam line versatility. We configured H6 to deliver a 120 GeV/c $\pi^-$ beam with maximum intensity $\approx 4\times10^{6}$ particles/spill. In this setup, the average energy deposited by beam particles in the detector central cell, determined from a simplified FLUKA~\cite{Battistoni:2015epi,Ahdida:2022gjl}-based simulation of the setup, is approximately 10~GeV; a rough estimate for the cut-off frequency reads $f_{cut}\approx 60$~MHz, having $N^\prime \approx 50 \times 10^3$~phe. The typical energy deposition per unit time $dE/dt$, at the maximum beam intensity, is approximately 8 GeV/$\mu$s. This is comparable to that expected for the full PKR-CAL, in which the 100 GeV/c $e^+$ beam deposits on average 68 GeV in the central cell, with maximum intensity of $10^5$ particles/s, resulting to $dE/dt=$7~GeV/$\mu$s. We exposed the POKERINO central cell to the beam, changing the opening of the various beamline collimators to vary the beam intensity, that was measured from the counting rate of a large plastic scintillator counter installed on the beamline upstream to our setup\footnote{Counter \texttt{XSCI 041.488}.}. In each run, we assessed the central cell gain by isolating events with small total energy deposition in the lateral crystals, imposing a 300 MeV threshold on the latter. For selected events, we measured the central cell energy distribution. This showed a clear peak for MIP-like events, in which the impinging hadron passes through the PbWO$_4$ crystals without any hard interaction. 
The position of the MIP peak $E_{MIP}$ was determined via an unbinned maximum-likelihood fit using a Landau distribution convoluted with a Gaussian curve. 
We assigned the RMS of the spill-by-spill $E_{MIP}$ distribution as the systematic uncertainty associated, for each run, with this observable. %

The correlation between $E_{MIP}$ and the beam intensity is shown in Fig.~\ref{fig:freq}, right panel. In case of multiple runs executed at the same intensity, we decided to report all data points in the graph. The superimposed red curve represents the model prediction with $f_{cut}=60$~MHz. This exhibit a non-negligible discrepancy with the experimental data, showing a stronger dependency on the frequency, down to a gain reduction of $\approx 3.5\%$ for the largest beam intensity. We attributed this discrepancy to other long-term effects such as the  decrease of the $\rm PbWO_4$ transparency due to the radiation damage induced by impinging
hadrons.  In order to check this hypothesis, we analysed data collected during two low-intensity
muon runs taken just before and just after the intensity scan, observing a decrease in the MIP peak position of approximately 4.5 \%. This value is comparable to observed $E_{MIP}$ reduction observed in the intensity scan. As such, we did not attempt a fit to the data to extract a quantitative estimate $f_{MIP}$, but we just observe that the data trend is qualitatively consistent with the expectations from the simplified model discussed above, with a modest decrease  of the MIP peak position at the highest beam intensities, slightly larger than that predicted from the value of $f_{cut}$ for this configuration. A more precise evaluation of $f_{MIP}$ in realistic beam conditions is postponed to the future PKR-CAL measurement, where a dedicated laser calibration system, currently in preparation, will allow to quantify precisely the radiation-induced  transparency degradation of $\rm PbWO_4$.


\section{Discussion and conclusions}

\section{Discussion and conclusions}

We characterized POKERINO, a small-scale prototype of the PKR-CAL electromagnetic calorimeter for a missing energy measurement within the NA64-$e$ experimental setup at the H4 beamline. The performed tests, including measurements on beam, with cosmsic rays and with a dedicated laser setup, confirmed that the adopted technical choices are compatible with the demanding performance requirements of the PKR-CAL; in particular, the light readout system based on SiPMs, which have never been used in high-energy electromagnetic calorimetry based on a fully homogeneous detector, performed as expected in the design phase, allowing to achieve the target energy resolution. The saturation effects due to the finite cell number of the SiPMs were studied in detail by exposing POKERINO to beams of different energies, and an optimized correction was devised so to mitigate this effect over all the energy range of the planned PKR-CAL measurements. 

The expected beam intensity represents a critical aspect for the operation of SiPMs at H4,  due to the negative-feedback mechanism induced by the  current through the SiPM bias resistors that can result in gain variations and, therefore, in the degradation of the energy resolution, especially in presence of instantaneous  intensity fluctuations of the beam. This effect was studied with dedicated tests aiming at reproducing the PKR-CAL measurement conditions, both by using a laser setup and by exposing POKERINO to high-intensity hadron beams. Overall, the obtained results indicate that the negative-feedback mechanism  does not represent a limiting factor for the energy resolution of POKERINO, (and hence for the PKR-CAL) at the nominal H4 beam intensities. Even in the presence of sizeable instantaneous beam intensity fluctuations, the expected contribution to the constant term of the calorimeter energy resolution is well below the dominant systematic effects. The agreement between the laboratory laser measurements, the simplified analytical model, and the preliminary beam test with POKERINO provides confidence that this effect can be reliably controlled in the full PKR-CAL detector, especially when combined with spill-by-spill calibration strategies based on high-energy electron events.

In conclusion, the performed measurements demonstrate the validity of the PKR-CAL technical choices,  proving the possibility to operate SiPMs in the demanding experimental conditions of a missing-energy measurement with a high-energy positron beam, and paving the way for the full PKR-CAL measurement with NA64-$e$ at CERN.  
\vspace{6pt} 

\section*{Author Contributions}
All authors have read and agreed to the published version of the manuscript.

\section*{Funding}
This work is part of a project that has received funding from the
European Research Council (ERC) under the European Union’s Horizon 2020 research and innovation programme, Grant agreement No. 947715 (POKER). 
This project has received funding from the European Union’s Horizon Europe Research and Innovation programme under Grant Agreement No 101057511 (EURO-LABS).

\section*{Data Availability}
Data will be made available on request.

\section*{Acknowledgments}
We gratefully acknowledge the technical staff of INFN, Sezione di Genova - F. Parodi, F. Pratolongo, E. Tixi - for their support during the construction and commissioning of the POKERINO detector. 
We acknowledge the support of the CERN management of staff for their vital contributions.
During the preparation of this manuscript, the authors used ChatGTP v. 5.2 for the purposes of cleaning the picture shown in Fig.~\ref{fig:POKERINO_setup_scheme_2024} from an object originally present in the background, irrelevant for the measurement. 
The authors have reviewed and edited the output and take full responsibility for the content of this publication.

\bibliographystyle{unsrt}  
\bibliography{biblio}      

\end{multicols}

\end{document}